\documentclass{aa}  

\usepackage{graphicx}
\usepackage{caption}
\usepackage{subcaption}
\usepackage{txfonts}
\usepackage{xcolor}
\usepackage{hyperref}

\newcommand{\nv}{N\,{\sc v}~}

\newcommand{\siiv}{Si\,{\sc iv}~}

\newcommand{\civ}{C\,{\sc iv}~}

\newcommand{\oiii}{O\,{\sc iii]}~}
\newcommand{\aliii}{Al\,{\sc iii}~}

\newcommand{\mgii}{Mg\,{\sc ii}~}
\newcommand{\feii}{Fe\,{\sc ii}~}
\newcommand{\feiii}{Fe\,{\sc iii}~}

\def\oiii{{\it{[OIII]}}}
\def\weak{{\it{Weak [OIII]}}}
\def\ltsima{$\; \buildrel < \over \sim \;$}
\def\simlt{\lower.5ex\hbox{\ltsima}}
\def\gtsima{$\; \buildrel > \over \sim \;$}
\def\simgt{\lower.5ex\hbox{\gtsima}}

\begin{document} 

   \title{The WISSH quasars project}
   \subtitle{VI. Fraction and properties of BAL quasars in the hyper-luminosity regime}
   \titlerunning{The BAL QSOs fraction and properties in WISSH} 

   \author{G. Bruni\inst{1}
   		\and E. Piconcelli \inst{2}
   		\and T. Misawa \inst{3}
		\and L. Zappacosta \inst{2}
		\and F. G. Saturni \inst{2,4}
		\and G. Vietri \inst{5,6}
		\and C. Vignali \inst{7,8}
		\and A. Bongiorno \inst{2}
		\and F. Duras \inst{9,2}
		\and C. Feruglio \inst{10,2}
		\and F. Tombesi \inst{11,12,13,2}
		\and F. Fiore \inst{10}
         	 }
   \institute{
        INAF - Istituto di Astrofisica e Planetologia Spaziali, via Fosso del Cavaliere 100, 00133 Roma, Italy\\ \email{gabriele.bruni@inaf.it}     
        \and INAF - Osservatorio Astronomico di Roma, via Frascati 33, 00040 Monte Porzio Catone, Roma, Italy
	\and School of General Education, Shinshu University 3-1-1 Asahi, Matsumoto, 390-8621, Japan
	\and ASI - Space Science Data Center, via del Politecnico snc, 00133 Roma, Italy
	\and Excellence Cluster Universe, Technische Universit\"at M\"unchen, Boltzmannstr. 2, 85748 Garching, Germany	
	\and  European Southern Observatory, Karl-Schwarzschild-Str. 2, 85748 Garching bei M\"unchen, Germany
	\and DiFA - Dipartimento di Fisica e Astronomia, Università degli Studi di Bologna, Via Gobetti 93/2, 40129 Bologna, Italy
	\and INAF - Osservatorio di Astrofisica e Scienza dello Spazio di Bologna, Via Piero Gobetti 93/3, 40129 Bologna, Italy
	\and Dipartimento di Matematica e Fisica, Università degli Studi Roma Tre, Via della Vasca Navale 84, 00146 Roma, Italy
	\and  INAF - Osservatorio Astronomico di Trieste, Via G.B. Tiepolo 11, 34143 Trieste, Italy
	\and Dipartimento di Fisica, Università degli Studi di Roma “Tor Vergata”, Via della Ricerca Scientifica 1, 00133 Roma, Italy
	\and Department of Astronomy, University of Maryland, College Park, MD 20742 USA
	\and NASA/Goddard Space Flight Center, Code 662, Greenbelt, MD 20771 USA		
	          }

   \date{}
   

 
  \abstract
   {The WISSH quasars project aims at studying the nuclear and host galaxy properties of the most luminous quasars ($L_{bol}>10^{47}$ erg/s, $1.8<z<4.6$), with special emphasis on the occurrence and physical parameters of winds at different scales.}
   {Nuclear winds are manifested as UV broad ($\geq$ 2,000 km/s) absorption lines (BAL) in $\sim$ 15\% of quasars. We aim at studying the incidence and properties of such winds in the WISSH sample, to investigate possible differences with respect to lower luminosity AGN regimes.}
   {We collected optical spectra from the SDSS data release 12, and identified those showing absorption troughs in the region between \siiv and \civ emission lines. We used three different indices for BAL absorption: the classic Balnicity Index (BI), the Absorption Index (AI), and the intermediate AI$_{1000}$.}
   {We find a higher observed fraction of \civ BAL quasars in the WISSH sample (24\%) with respect to previous catalogues (10-15\%). These WISSH BAL quasars are also characterized by a larger average BI ($\sim$4,000 km/s) and maximum velocity ($\sim$17,000 km/s). Moreover, for two objects we discovered BAL features bluewards of the \siiv peak, which can be associated to \civ absorption with velocity of 0.15c. Finally, we updated previous studies on the dependence of maximum outflow velocity upon bolometric luminosity, showing that BAL winds have intermediate properties compared to molecular/ionized winds and ultra fast outflows (UFOs). Finally, the radio properties of the WISSH BAL quasars as a whole are in line with those of samples at lower luminosities from previous studies.}
   {Our results suggest that the higher $L_{\rm bol}$ of the WISSH quasars likely favours the acceleration of BAL outflows  and that their most likely driving mechanism is radiation pressure. Furthermore we estimate that the kinetic power associated to these winds in hyperluminous quasars is sufficient, for the highest column density and fastest winds, to provide efficient feedback onto the host galaxy.}
   
   \keywords{galaxies:active; galaxies: nuclei - quasars: absorption lines; quasars:general; quasars: supermassive black holes - ISM: jets and outflow}

   \maketitle


\section{Introduction}

In the last two decades, there has been growing recognition of the potential importance of quasar (QSOs) winds for the growth of super-massive black holes (SMBH, \citealt{1998A&A...331L...1S}), enrichment of the intergalactic medium (\citealt{2007ApJ...665..187L}), galaxy formation (\citealt{2006ApJ...650....7H}), evolution of the host galaxy (\citealt{2005Natur.433..604D}), and the luminosity function of quasars (\citealt{2003ApJ...595..614W}). 
The Active Galactic Nucleus (AGN) power is fundamental in the overall winds dynamics, thus bolometric luminosity ($L_{bol}$) and Eddington ratio are key quantities to understand the feedback mechanism on the host galaxy. Interaction of radiation, jets, and winds with the interstellar medium of the host is often invoked as responsible of star-formation quenching and growth regulation, and commonly explained with two main scenarios: the radio-mode, and the quasar-mode one (see \citealt{2012ARA&A..50..455F} for a review). In the former, the collimated radio jet is responsible for removing - through fast shocks - the ambient gas, while in the latter the main cause is identified in fast, nuclear winds. Recently, both scenarios have proven to be valid: radio-mode feedback has been studied by \cite{2014Natur.511..440T}, who presented evidences of a molecular outflow accelerated by the jet in a Seyfert galaxy, while \cite{2015Natur.519..436T} and \cite{2015A&A...583A..99F} pointed out that highly-ionized, massive SMBH winds with velocity of 0.1-0.2$c$ are likely the drivers of kpc-scale molecular winds with large outflow rates, i.e. $\gtrsim$500 M$_\odot$/yr (see \citealt{2019arXiv190310528B} for a more complete picture).

Nuclear winds from the accretion disk can be observed in the UV domain as broad absorption lines (BALs) blue-wards of prominent emission lines (e.g. \civ, \siiv) in 10-20\% of optically-selected quasars. They trace wind velocities from a few thousands km/s up to $\sim$0.3$c$ (e.g. \citealt{2003AJ....125.1784H, 2018MNRAS.476..943H}). 
Depending on the involved species, BALs are divided into high-ionization (HiBALs) and low-ionization (LoBALs): while high-ionization species (\civ, \siiv, \nv) always produce the most prominent absorption features in these objects, about 15\% shows also troughs blue-wards of lower-ionization species like \mgii and \aliii. Additionally, are termed FeLoBALs the LoBALs showing also \feii and \feiii absorption features.  The number of known BAL QSOs has greatly increased in recent years, thanks to the several data releases of the Sloan Digital Sky Survey (SDSS, \citealt{2000AJ....120.1579Y}), reaching more than 20,000 objects in the recent releases of the SDSS quasar catalogue \citep{2017A&A...597A..79P, 2018A&A...613A..51P}. This allowed detailed studies of statistically complete samples, helping in characterizing the BAL phenomenology (see Sec. 3 for a complete discussion). 
Notwithstanding, a comprehensive scheme for the launching and geometry of BAL winds is still missing, although some attempts of constructing reasonable scenarios exist (\citealt{1995ApJ...454L.105M, 2000ApJ...538..684P, 2000ApJ...545...63E, 2010A&A...516A..89R}). Despite theoretical models suggest that these winds should be launched by the accretion disk, at a radius smaller than 1 pc (e.g. \citealt{2000ApJ...538..684P, 2000ApJ...545...63E}), to date observational results suggest that the distance at which the absorption occurs ranges from several tens to thousands parsecs, with larger radii for luminous quasars (see e.g. 
\citealt{2013MNRAS.436.3286A,2019ApJ...872...21H} and references therein). Lately,
\cite{2018ApJ...857...60A} found evidence that 50\% of BAL winds extend at least 100 pc from the nucleus, in a sample of $\sim$20 high-luminosity quasars. This highlights that these ionized outflows could indeed represent  an important source of feedback to the host galaxy, although this is not in line with previous claims that BAL winds are located at scales of the AGN accretion disk.

Similarly, Ultra Fast Outflows (UFOs) are of nuclear origin as well but detected as absorption troughs in  the X-ray domain, tracing higher-ionization species at mildly relativistic velocities $>0.1c$ (\citealt{2010A&A...521A..57T, 2011ApJ...742...44T}). The origin of both BAL and UFO winds should reside in the accretion disk, and be driven by radiation pressure or magnetohydrodynamic processes (\citealt{2002ApJ...569..641L, 2013MNRAS.430.1102T, 2018ApJ...852...35K}). Radio emission from the jet was used by different authors as indicator of the outflow orientation in samples of radio-loud BAL QSOs, not finding a clear hint of a preferred one (\citealt{2008MNRAS.388.1853M, 2011ApJ...743...71D, 2012A&A...542A..13B}). VLBI observations showed a variety of morphologies (\citealt{2013A&A...554A..94B}) also not pointing towards a particular angle, suggesting that the presence of BAL outflows and the consequent fraction of these objects among quasars should be due to inner physical properties of the AGN, and not to a mere orientation effect. Indeed, indications of an anticorrelation with radio loudness has been found (\citealt{2000ApJ...538...72B, 2006ApJ...641..210G, 2008ApJ...687..859S}), while \cite{2003AJ....125.1784H} suggested that optically-bright BAL QSOs are half as likely as non-BAL QSOs to have $S_{1.4 \rm{GHz}}>1$ mJy.

AGN winds at different scales are expected to increase their momentum and kinetic power with AGN bolometric luminosity (\citealt{2008ApJ...686..219M, 2012ApJ...745L..34Z, 2014MNRAS.444.2355C}). There is mounting evidence that this prediction is correct based on observations at different wavelength of  outflows involving different gas phases and distances from the SMBH (e.g., \citealt{2014A&A...562A..21C, 2017MNRAS.472L..15M, 2017A&A...601A.143F}). We have therefore undertaken a multi-band (from millimeter to X-rays) follow-up of a sample of 86 WISE/SDSS selected hyper-luminous (WISSH) quasars in the redshift range  z$\approx$ $2-4$, i.e the so-called `cosmic noon' at the peak of star formation activity and QSO number density. The major goal is to provide a detailed investigation of  nuclear and host galaxy properties and the census of AGN-driven winds in sources at the brightest end of the AGN luminosity function. WISSH QSOs exhibit very high bolometric luminosity ($L_{bol}\gtrsim10^{47}$ erg/s) powered by highly-accreting, ultra-massive ($>$ 10$^{9}$ M$_\odot$) SMBHs (see \citealt{2017A&A...598A.122B, 2017A&A...604A..67D, 2017MNRAS.468.3150M, 2018A&A...617A..81V}). These hyper-luminous AGN are thus expected to launch the most powerful outflows. By analyzing rest-frame UV and optical spectra of WISSH QSOs, \cite{2017A&A...598A.122B, 2018A&A...617A..81V} indeed reported on the discovery of ionized outflows both on kpc- and pc-scale, with extreme properties in terms of velocity and kinetic energy. 

In this paper, we present a study of the fraction and properties of the BAL QSOs in the WISSH sample, discussing their dependence on the extreme bolometric luminosities of these objects. We adopt the latest cosmological parameters from the \emph{Planck} mission (\citealt{2018arXiv180706209P}), i.e. assuming the base-$\Lambda\rm{CDM}$ cosmology: $H_{0}=67.4$ km/s/Mpc, $\Omega_{\rm{m}}=0.315$, and $\Omega_{\Lambda}=0.685$.


\begin{table*}
 \centering
   \caption{The 40 BAL QSOs from the WISSH sample: the first 38 objects are identified through absorption features between the \siiv and \civ emission lines, while the last two ones from features blue-wards of \siiv. Absorption indices estimates are given in columns 4-6. In the last column, the BAL type is reported: objects with a \civ BI$>$0 are in bold face, while "BAL" means that the spectrum coverage does not allow a proper HiBAL/LoBAL classification. Starred redshifts are from \cite{2018A&A...617A..81V}.}
  \label{sources}
 \begin{tabular}{cclrrrrrc}
\hline
  \multicolumn{1}{c}{ID} 				&
  \multicolumn{1}{c}{SDSS ID} 			&
  \multicolumn{1}{c}{$z$} 				&
  \multicolumn{1}{c}{AI} 				&
  \multicolumn{1}{c}{AI$_{1000}$} 		&
  \multicolumn{1}{c}{BI} 				&
  \multicolumn{1}{c}{$v_{min}$} 		&
  \multicolumn{1}{c}{$v_{max}$} 		&
  \multicolumn{1}{c}{Type} 				\\
  
   \multicolumn{1}{c}{}         	    &
   \multicolumn{1}{c}{}         	    &
   \multicolumn{1}{c}{}             	&  
   \multicolumn{1}{c}{[km/s]}           &
   \multicolumn{1}{c}{[km/s]}           &
   \multicolumn{1}{c}{[km/s]}      	    &  
   \multicolumn{1}{c}{[km/s]}           &
   \multicolumn{1}{c}{[km/s]}           &
   \multicolumn{1}{c}{}         	    \\    
       
   \multicolumn{1}{c}{(1)}         	    &
   \multicolumn{1}{c}{(2)}          	&
   \multicolumn{1}{c}{(3)}        	    &
   \multicolumn{1}{c}{(4)}          	&
   \multicolumn{1}{c}{(5)}          	&
   \multicolumn{1}{c}{(6)}          	&
   \multicolumn{1}{c}{(7)}          	&
   \multicolumn{1}{c}{(8)}          	&
   \multicolumn{1}{c}{(9)}          	\\
\hline
  0045+14   	& SDSS004527.68+143816.1    	& 1.992  			& 7762  	& 7718  	& 7016  	& 1300  	& 16000  	& {\bf LoBAL}   \\
  0216$-$09 	& SDSS021646.94$-$092107.2  	& 3.691  			& 2335  	& 1940  	& 1170  	& 8770  	& 18800  	& {\bf HiBAL}   \\
  0414+06   	& SDSS041420.90+060914.2    	& 2.614$^\star$  	& 10806 	& 10285 	& 7415  	& 1140  	& 13750  	& {\bf HiBAL }  \\
  0747+27   	& SDSS074711.14+273903.3    	& 4.110  			& 209   	& 12    	& 0     	& 2940  	& 12680  	& BAL           \\
  0928+53   	& SDSS092819.29+534024.1    	& 4.390  			& 9217  	& 8327  	& 8327  	& 3430  	& 19050  	& {\bf BAL}     \\
  0959+13   	& SDSS095937.11+131215.4    	& 4.061  			& 29    	& 0     	& 0     	& 13800   	& 15700  	& BAL           \\
  1013+56   	& SDSS101336.37+561536.3    	& 3.633  			& 127   	& 0     	& 0     	& 3460  	& 11650  	& HiBAL         \\
  1025+24   	& SDSS102541.78+245424.2    	& 2.382  			& 3059  	& 2866  	& 2356  	& 7260  	& 13750  	& {\bf LoBAL}   \\
  1048+44   	& SDSS104846.63+440710.8    	& 4.347  			& 11125 	& 10720 	& 9044  	& 2060  	& 17840  	& {\bf BAL}     \\
  1051+31   	& SDSS105122.46+310749.3    	& 4.243  			& 203   	& 14    	& 0     	& 9920  	& 11800  	& BAL           \\
  1103+10   	& SDSS110352.74+100403.1    	& 3.590  			& 357   	& 38    	& 0     	& 1530  	& 6160   	& HiBAL         \\
  1110+19   	& SDSS111017.13+193012.5    	& 2.498  			& 2680  	& 2532  	& 0     	& 590   	& 4900   	& HiBAL         \\
  1110+48   	& SDSS111038.63+483115.6    	& 2.957  			& 145   	& 0     	& 0     	& 14360 	& 15440  	& HiBAL         \\
  1122+16   	& SDSS112258.77+164540.3   	 	& 3.024  			& 8061  	& 7861  	& 7312  	& 6610  	& 17670  	& {\bf LoBAL}   \\
  1130+07   	& SDSS113017.37+073212.9    	& 2.654  			& 25    	& 0     	& 0     	& 10710 	& 12140  	& HiBAL         \\
  1157+27   	& SDSS115747.99+272459.6    	& 2.217$^\star$  	& 5641  	& 4948  	& 4021  	& 2000  	& 17200  	& {\bf HiBAL}   \\
  1204+33   	& SDSS120447.15+330938.7    	& 3.596  			& 5282  	& 5117  	& 3531  	& 1620  	& 10590  	& {\bf LoBAL}   \\
  1210+17   	& SDSS121027.62+174108.9    	& 3.604$^\star$  	& 5903  	& 5595  	& 5116  	& 5840  	& 14750  	& {\bf HiBAL}   \\
  1215$-$00 	& SDSS121549.81$-$003432.1  	& 2.707  			& 1679  	& 1289  	& 399   	& 5800  	& 20640  	& {\bf HiBAL}   \\
  1237+06   	& SDSS123714.60+064759.5    	& 2.781  			& 872   	& 187   	& 0     	& 4200  	& 19810  	& HiBAL         \\
  1245+01   	& SDSS124551.44+010505.0    	& 2.798  			& 3707  	& 3530  	& 1939  	& 1300  	& 12300   	& {\bf HiBAL}   \\
  1250+20   	& SDSS125050.88+204658.7    	& 3.570  			& 890   	& 824   	& 0     	& 350   	& 3820   	& HiBAL         \\
  1305+05   	& SDSS130502.28+052151.1    	& 4.071  			& 246   	& 42    	& 0     	& 19500  	& 22550  	& HiBAL         \\
  1326$-$00 	& SDSS132654.96$-$000530.1  	& 3.303$^\star$  	& 2275  	& 1960  	& 0     	& 0  		& 3850   	& HiBAL         \\
  1328+58   	& SDSS132827.06+581836.8    	& 3.133  			& 322   	& 131   	& 0     	& 3500   	& 5400   	& HiBAL         \\
  1422+44   	& SDSS142243.02+441721.2    	& 3.648$^\star$  	& 310   	& 0     	& 0     	& 15410 	& 22780  	& HiBAL         \\
  1447+10   	& SDSS144709.24+103824.5    	& 3.699  			& 6179  	& 4484  	& 65    	& 370   	& 23520  	& {\bf HiBAL}   \\
  1451+14   	& SDSS145125.31+144136.0    	& 3.102  			& 4848  	& 4340  	& 3586  	& 6000  	& 13400  	& {\bf HiBAL}   \\
  1506+52   	& SDSS150654.55+522004.7    	& 4.068  			& 5147  	& 4938  	& 4396  	& 12970 	& 23130  	& {\bf HiBAL}   \\
  1513+08   	& SDSS151352.52+085555.7    	& 2.897  			& 1293  	& 820   	& 0     	& 200   	& 4800  	& HiBAL         \\
  1544+41   	& SDSS154446.34+412035.7    	& 3.548  			& 4808  	& 4663  	& 4270  	& 9580  	& 23500  	& {\bf HiBAL}   \\
  1549+12   	& SDSS154938.72+124509.1    	& 2.365$^\star$  	& 4918  	& 4425  	& 1735  	& 0     	& 6470   	& {\bf HiBAL }  \\
  1555+10   	& SDSS155514.85+100351.3    	& 3.512  			& 6992  	& 6490  	& 5576  	& 7090  	& 16000  	& {\bf LoBAL}   \\
  1633+36   	& SDSS163300.13+362904.8    	& 3.576  			& 377   	& 241   	& 90    	& 18700  	& 24800  	& {\bf HiBAL }  \\
  1639+28   	& SDSS163909.10+282447.1    	& 3.801  			& 366   	& 259   	& 0     	& 22300 	& 28460  	& HiBAL         \\
  1650+25   	& SDSS165053.78+250755.4    	& 3.338  			& 3675  	& 3516  	& 2913  	& 5000  	& 13100  	& {\bf HiBAL}   \\
  2123$-$00 	& SDSS212329.46$-$005052.9 	    & 2.282$^\star$  	& 107   	& 0     	& 0     	& 13500  	& 24600  	& HiBAL         \\
  2238$-$08 	& SDSS223808.07$-$080842.1 	    & 3.122  			& 4300  	& 3889  	& 2228  	& 1500  	& 19000  	& {\bf HiBAL}   \\
  \hline
  0947+14   	& SDSS094734.19+142116.9    	& 3.040   			& 795   	& 685   	& 548   	&  41250	& 46040 	& HiBAL         \\
  1538+08   	& SDSS153830.55+085517.0    	& 3.567$^\star$ 	& 439   	& 318   	& 130   	&  38300 	& 47000   	& HiBAL         \\
\hline
\end{tabular}
\end{table*}


\section{BAL identification from SDSS DR12 spectra}

We performed a detailed search of BAL QSOs in the WISSH sample of 86 hyper-luminous QSOs, making use of SDSS optical spectra. At the mean redshift of the sample ($z\sim$3.2) the SDSS observer's frame wavelength range (3500-9000 \AA) corresponds to 830-2140 \AA\, in the rest-frame, allowing to explore the whole region between Ly$\alpha$ and \aliii emission lines.
Our aim was to test whether the BAL fraction, and thus the presence of nuclear winds, is different at high-luminosity regime, comparing with previous estimates from the literature at lower luminosity ($L_{bol}<10^{47}$ erg/s). In the following, the method and details for BAL QSOs selection are given.

We collected optical spectra of all the 86 WISSH objects from the 12$^{th}$ data release of the SDSS (\citealt{2015ApJS..219...12A}). As a first selection step, we visually inspected them to find footprint of absorption features blue-wards of the \civ emission peak. This led us to 42 candidate BAL QSOs ($\sim$48\% of the sample). Then, we fitted their spectra with the {\tt{continuum}} tool in {\tt{IRAF}}\footnote{\href{http://iraf.noao.edu/}{http://iraf.noao.edu/}}, using {\tt{spline3}} as polynomial function, and flagging all the points part of the \civ absorption feature. With the obtained residuals, we characterized in velocity space the broad absorption feature making use of three well-known indices from the literature:
1) the Absorption Index (AI), as defined in \cite{2002ApJS..141..267H},
\begin{equation}
{\rm{AI}}=\int_{0}^{25000}(1-\frac{f(v)}{0.9})\cdot Cdv,
\end{equation}
where the parameter $C$ is unity over contiguous troughs of at least 450 km/s;
2) the modified Absorption Index (AI$_{1000}$, \citealt{2012A&A...542A..13B}), defined as in \cite{2002ApJS..141..267H}, but where the parameter $C$ is unity over contiguous troughs of at least 1,000 km/s (as in \citealt{2006ApJS..165....1T}); 
3) the Balnicity Index (BI), as defined by \cite{1991ApJ...373...23W},
\begin{equation}
{\rm{BI}}=\int_{3000}^{25000}(1-\frac{f(v)}{0.9})\cdot Cdv,
\end{equation}
where the parameter $C$ is unity over contiguous troughs of at least 2,000 km/s. To perform this calculation, the spectral region between the peaks of the \civ and \siiv emission lines was integrated up to 25,000 km/s from the former, starting from a minimum detachment of 0 km/s for AI and AI$_{1000}$, while 3,000 km/s for BI. The three indices are increasingly conservative, from AI to BI, AI$_{1000}$ being the one that allows to study a variety of absorption features, but still filtering the most ambiguous ones. Although they can be considered as velocity-weighted equivalent widths, they do not directly measure any outflow physical quantity, but instead can be used to quantify the strength and width of the absorption for BAL QSOs identification purposes.
Two among the objects presented here (0414+06 and 1210+17) are newly found BAL QSOs, since not present in previous BAL catalogues (\citealt{2006ApJS..165....1T, 2009ApJ...692..758G, 2011MNRAS.410..860A}) or flagged as BAL in the latest editions of the SDSS QSOs catalogue (\citealt{2014A&A...563A..54P, 2017A&A...597A..79P} - the latter extracted from SDSS DR12 as for our sample). Both objects show issues in the spectrum, that most probably misled the algorithm used for the compilation of previous catalogues: 0414+06 presents a spike in correspondence of the \civ peak, while 1210+17 has a spike at the right edge of the spectrum, and a very faint Ly-$\alpha$ line, that does not allow to easily identify the \siiv and \civ emission lines. For both objects a wrong redshift is given in SDSS, so we provide the corrected estimate in Tab. \ref{sources}. In total, we found in WISSH 38 objects with an AI$>$0 (44$\pm$7\%), 32 with AI$_{1000}>$0 (37$\pm$6\%), and 21 with BI$>$0 (24$\pm$5\%), whose uncertainties are calculated with Poissonian statistic. Spectra for all the 38 BAL QSOs are given in appendix.

In addition to the classic selection based on \civ, we also looked for additional absorption from other species known to show BAL features, namely \siiv and \aliii (\mgii was not covered by the spectra), with the aim to perform a HiBAL/LoBAL classification. In order to quantify the absorption for these two species we used a modified version of the previous indices, shifting the integration range blue-wards of the \siiv and \aliii emission lines, respectively, and adopting the same maximum velocity of 25,000 km/s. Two objects show strong BAL features blue-wards of the \siiv emission line, while only narrow absorption in the range between \siiv and \civ: we discuss these two particular cases in section \ref{Si_BALs}. For 5 objects the redshift did not allow us to have the \aliii emission line region in the SDSS wavelength range, so the HiBAL/LoBAL classification was not possible. For all of the other sources the classification is given in Tab. \ref{sources}.


\begin{figure}
\centering
\begin{subfigure}[a]{1.0\linewidth}
\includegraphics[width=\linewidth]{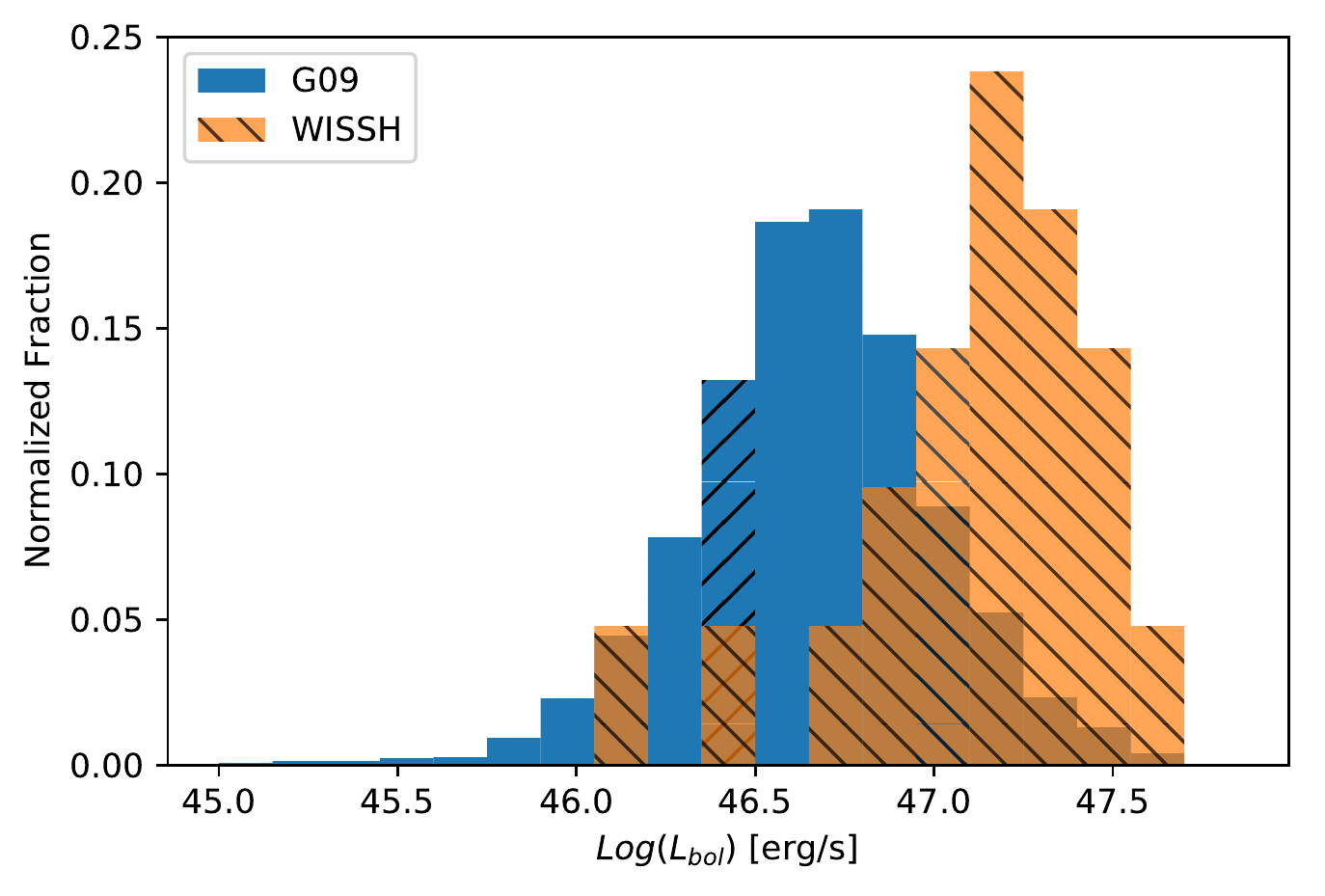}
\end{subfigure}
\begin{subfigure}[b]{1.0\linewidth}
\includegraphics[width=\linewidth]{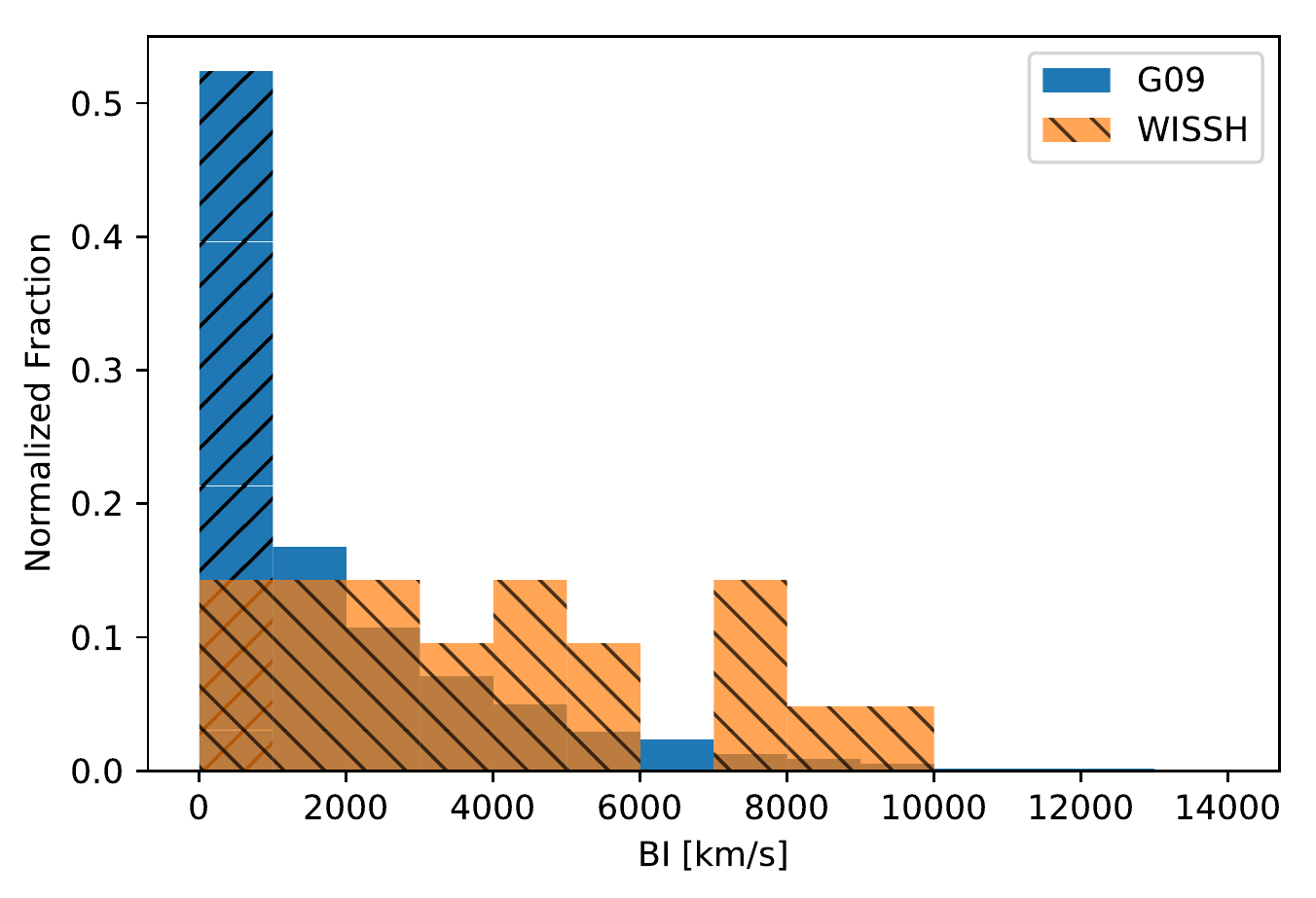}
\end{subfigure}
\begin{subfigure}[c]{1.0\linewidth}
\includegraphics[width=\linewidth]{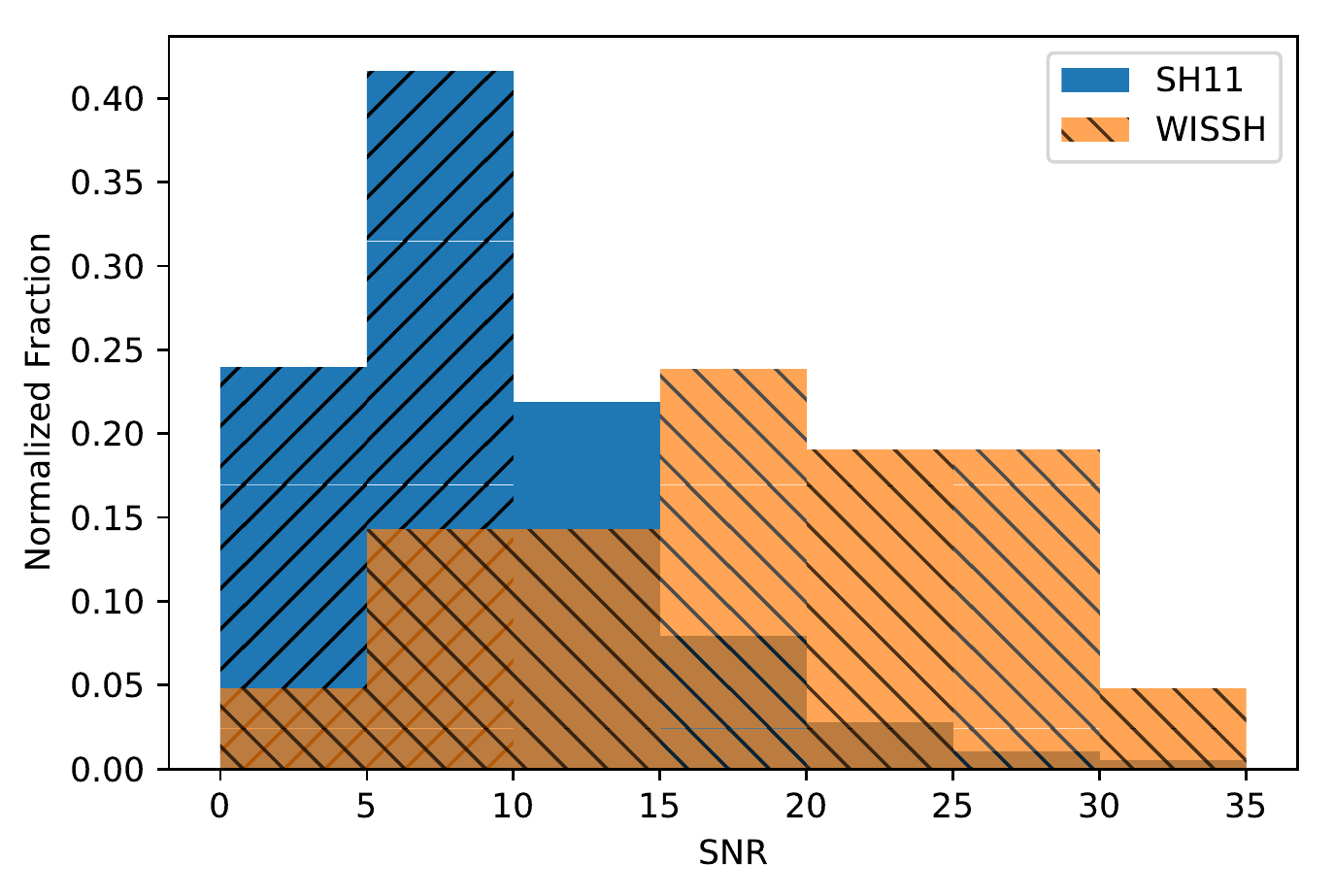}
\end{subfigure}

\caption{
Top panel: bolometric luminosity normalized distributions for WISSH and G09 objects with \civ BI$>$0; 
middle panel: \civ BI distribution for the WISSH sample and the BAL QSOs catalogue from \cite{2009ApJ...692..758G}; 
bottom panel: normalized distributions for the 1500-1600 \AA\ SNR values extracted from SDSS DR7, for \civ BI$>$0 BAL QSOs in WISSH and in the \cite{2011ApJS..194...45S} QSO catalogue (SH11).} \label{BI+SNR}
\end{figure}


\section{BAL fraction and strength at the extreme luminosity regime}

\subsection{The BAL fraction in WISSH}
The observed fraction of BAL QSOs ($\mathcal{F}_{obs}$) in the total AGN population has been explored by several authors since the early 2000s. The main works trying to compile BAL QSOs catalogues and study their characteristics are the following: 1) \cite{2003AJ....125.1784H}, presenting a sample of 67 BAL QSOs from the large bright quasar survey (\citealt{1995AJ....109.1498H}, pre-SDSS era), resulting in a $\mathcal{F}_{obs}$=15$\pm$3\%; 2) \cite{2003AJ....125.1711R}, producing the first catalogue from SDSS early data release, including more than 200 objects, giving $\mathcal{F}_{obs}$=14.0$\pm$1.0; 3) \cite{2006ApJS..165....1T}, from SDSS DR3, $\mathcal{F}_{obs}\sim$10.4\%; 4) \cite{2008MNRAS.386.1426K}, from SDSS DR3, $\mathcal{F}_{obs}$=13.5;  5) \cite{2009ApJ...692..758G} (G09 hereafter), from SDSS DR5 as well, $\mathcal{F}_{obs}$=13.3$\pm$0.6\%; 6) \cite{2011MNRAS.410..860A}, the latest one, from SDSS DR6, finding a dependence from redshift and a $\mathcal{F}_{obs}=8.0\pm0.1$\%. The common criterion adopted in the mentioned works to estimate $\mathcal{F}_{obs}$ is the \civ BI. Considering the same definition of BAL QSO, we find in WISSH an observed fraction $\mathcal{F}_{obs}^{\small{W}}$=24$\pm$5\% having a \civ BI$>$0: this is almost twice the average fraction found in previous works, and 9\% more than the largest one among them. This points towards a intrinsic difference among WISSH BAL QSOs and previous samples, possibly linked to the hyper-luminosity regime. 
\cite{2008ApJ...672..108D} investigated $\mathcal{F}_{obs}$ among 2MASS-selected QSOs: they claimed a value of $\sim40$\%, finding that BAL QSOs are redder than non-BAL QSOs, and suggesting that a negative selection bias prevents from correctly estimating $\mathcal{F}_{obs}$ in the optical band. Nevertheless, they found that a more restrictive classification based in BI leads to a $\mathcal{F}_{obs}\sim20-23$\% (see their Fig. 6), more simlar to the fractions found from optical band studies. This value is directly comparable, and in agreement, with the fraction we are finding for our sources. We also note that their 2MASS-selection (i.e. $K_s<15.1~\rm  mag$) for $z>1.7$ sources implies a selection of the most luminous QSOs with $M_{K_s}<-30~\rm mag$ in the optical rest-frame band\footnote{Notice that the centroid of the $K_s$ filter is centered in the rest-frame range $4000-8000\rm \AA$ for their sources at $z=1.7-4.38$.}. Indeed as stated by \cite{2008ApJ...672..108D}, only $\sim$5\% of SDSS QSOs are detected in 2MASS, as the latter is significantly shallower than the former. This suggests that the luminosity mainly drives the increase in the BAL fraction.

We could also estimate the fraction of LoBALs among BAL QSOs in WISSH. Indeed, for 19 objects among the 21 ones with BI$>$0 the spectrum covers the \aliii region, allowing the LoBAL classification: for 5 over 19 we obtained a \aliii BI$>$0, giving a LoBAL fraction of $\sim26^{+18}_{-11}$\% (following Poissonian statistics for small numbers, see \citealt{1986ApJ...303..336G}). Generally, LoBALs can be easily missed in large surveys, due to reddening and complex spectral features. The fraction found in WISSH is compatible within the errors with the one reported in the literature for SDSS objects ($\sim$15\%; e.g., \citealt{2003AJ....125.1711R, 1992ApJ...390...39S}). Nevertheless, considering all \civ BI$>$0 from G09, and applying the same criterium as above for LoBAL identification (i.e. \civ BI$>$0 and \aliii BI$>$0), we find a much lower LoBAL fraction of 6.5$\pm$0.4\%: this could indicate that at the WISSH hyper-luminosity regime not only highly-ionized gas outflows are more common, but also the lower-ionized component - possibly launched from larger accretion disk radii - can more easily reach the relativistic velocities needed to produce BAL troughs. 

In the rest of the paper we will consider the G09 catalogue as the comparison sample, since it provides a more complete collection of wind parameters to be compared with our sample. In addition to that, we consider some quantities estimated in the \cite{2011ApJS..194...45S} QSOs catalogue, based on SDSS DR7, that adopt the BAL classification of G09. Considering all objects from G09 with a \civ BI$>$0, we have a comparison sample of 4,242 objects with redshift in the range 1.5$<z<$5.0. In order to study the BAL winds dependence on $L_{\rm bol}$, and compare the G09 sample with the hyper-luminous objects from WISSH, we cross-correlated the \civ BI$>$0 sample above with the \cite{2011ApJS..194...45S} QSOs catalogue (1 arcsec match in position) as this work provides the $L_{bol}$ estimates needed for our analysis. This reduces the number of comparison BAL QSO objects to 3874. Fig. \ref{BI+SNR} (top panel) shows the $L_{\rm bol}$ distributions for the \civ BI$>$0 objects from G09 and WISSH, and highlights the extreme values of $L_{\rm bol}$ covered by the WISSH QSOs with respect to the previous samples.

\subsection{BI distribution and BAL strength}
As a further comparison with the G09 BAL QSOs population, we investigated the BI distributions for all objects having a \civ BI$>$0 from G09 and from the WISSH sample: an excess of high BI values is found (see Fig. \ref{BI+SNR}). A Kolmogorov-Smirnov (KS) test gives a $p<0.05$ for the two distributions to be drawn from the same one. 

G09 found a dependence of the BAL fraction from the spectra Signal-to-Noise Ratio (SNR), with a higher SNR in the \civ region favoring the identification of BAL troughs, and thus enhancing the fraction of BAL QSOs for a given sample. The WISSH sample is composed by high-luminosity objects (see Fig. \ref{BI+SNR}), thus a higher SNR is expected. To verify how much this effect could influence the fraction we measured in WISSH, we extracted the SNR of the 1500-1600 \AA\, rest-frame, non-absorbed, region from the \cite{2011ApJS..194...45S} quasar catalogue for the objects with a BI$>$0 in WISSH, and for all the objects flagged as BAL (as noted before, the \citealt{2011ApJS..194...45S} BAL classification is taken from G09, and therefore it allows for a proper comparison). In Fig. \ref{BI+SNR}, we compare the SNR distribution for WISSH BI$>$0 objects with the one for all the BI$>$0 objects from \cite{2011ApJS..194...45S}. The mean SNR value increases in the two sets, from a value of 9 per pixel for the \cite{2011ApJS..194...45S} one, to 18 for WISSH. Following G09, an SNR increase from 9 to 18 should imply a variation of about 2\% more BAL QSOs. This is clearly a too low contribution to justify the higher BAL QSOs fraction we see in the WISSH sample with respect to the literature. 

\begin{figure}
\centering
\includegraphics[width=1.0\linewidth]{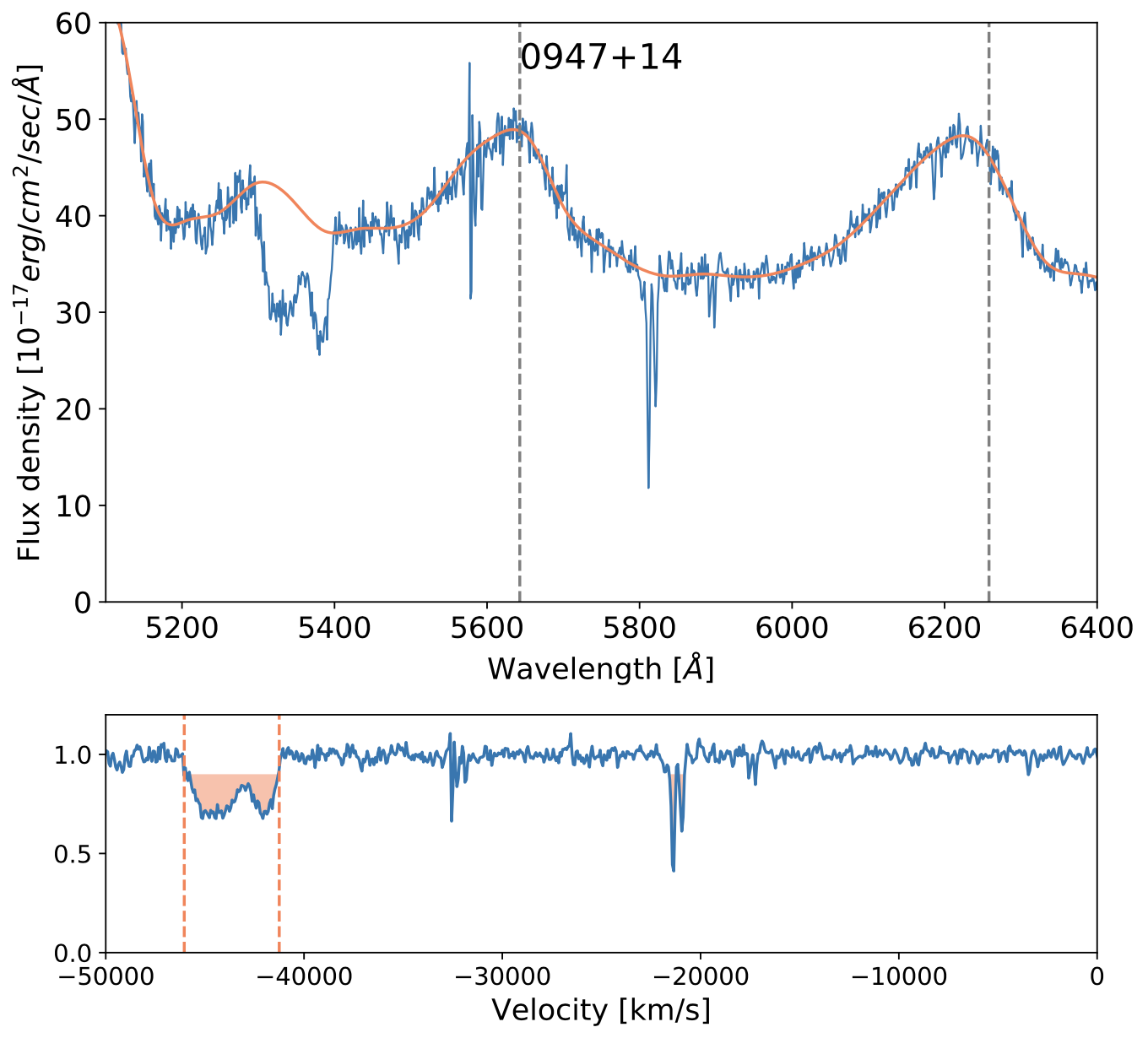}
\caption{Zoom on the \civ and \siiv regions for 0947+14 (SDSS DR12 spectrum), one of the two objects showing ultra-fast ($>$0.1c) BAL signatures associated to \civ absorption, blue-wards of the \siiv emission line. Top panel: spectrum (blue line) and fit performed in {\tt IRAF}; dashed lines indicate the position of the \siiv and \civ peaks as calculated from redshift. Bottom panel: residuals between the \civ peak and $-$50,000 km/s, with absorption below 90\% of the continuum highlighted in orange; dashed lines indicate the minimum and maximum velocity estimated for the BAL outflow.}
\label{Siiv}
\end{figure}

\subsection{Detection of Ultra-fast BAL Outflows (UBOs)}
\label{Si_BALs}
In addition to the 38 \civ BAL QSOs presented in the previous section, we could identify two more objects (0947+14 and 1538+08) showing BAL signatures blue-wards of the \siiv emission line, but only very narrow absorption troughs between the \civ and \siiv lines (see Fig. \ref{Siiv}). The optical depth ratio of the two species varies as a function of the ionisation parameter $U$, and for values typically found in BAL QSOs ($Log(U)>-2$) \civ dominates the \siiv depth by a factor of $\sim$2 (\citealt{2012ApJ...750..143D}). The absence of strong absorption between the \siiv and \civ lines can thus suggest that the BAL on the left side of \siiv is indeed due to \civ. This leads to a maximum velocity for the outflows in 0947+14 and 1538+08 of 0.15$c$ and 0.16$c$, respectively. These  ultra-fast BAL outflows (UBOs) - that we could conveniently define as the ones having a maximum velocity larger than 0.1$c$, allowing them to show \civ absorption blue-wards of the \siiv emission peak - are  among the most extreme ones detected to date. Although less common than slower BAL outflows, there is a growing census of UBOs in the literature \citep{10.1111/j.1365-2966.2010.17677.x, 2013MNRAS.435..133H, 2016MNRAS.457..405R}, presenting in some cases \civ absorption troughs bluewards of the Ly$\alpha$ peak, implying a velocity of $\sim$0.3$c$ (\citealt{2018MNRAS.476..943H}).
A detailed analysis of multi-epoch UBOs properties in the WISSH sample will be presented in a forthcoming paper (Piconcelli et al. in prep.). 

We explored the relative abundance of such relativistic BAL features in G09, by identifying all objects having a \siiv BI$>$0 and a \civ BI=0 in analogy to the two ones found in WISSH. In total, 2.9$\pm$0.3\% of objects in G09 show such characteristics (127 over the 4369 ones having \siiv or \civ BI$>$0), while the fraction we obtain for WISSH is $9^{+11}_{-6}$\% (2 over 23 objects, i.e. the 21 \civ BI$>$0 BAL QSOs plus the 2 UBOs themselves). This hints to a possible larger fraction of UBOs in high-luminosity QSOs such as WISSH. However, the very limited statistics hampers any firm conclusions on this trend and additional investigations based on larger samples of QSOs with $L_{bol}>10^{47}$ erg/s are necessary.



\begin{figure}
\centering

\begin{subfigure}[a]{1.0\linewidth}
\includegraphics[width=\linewidth]{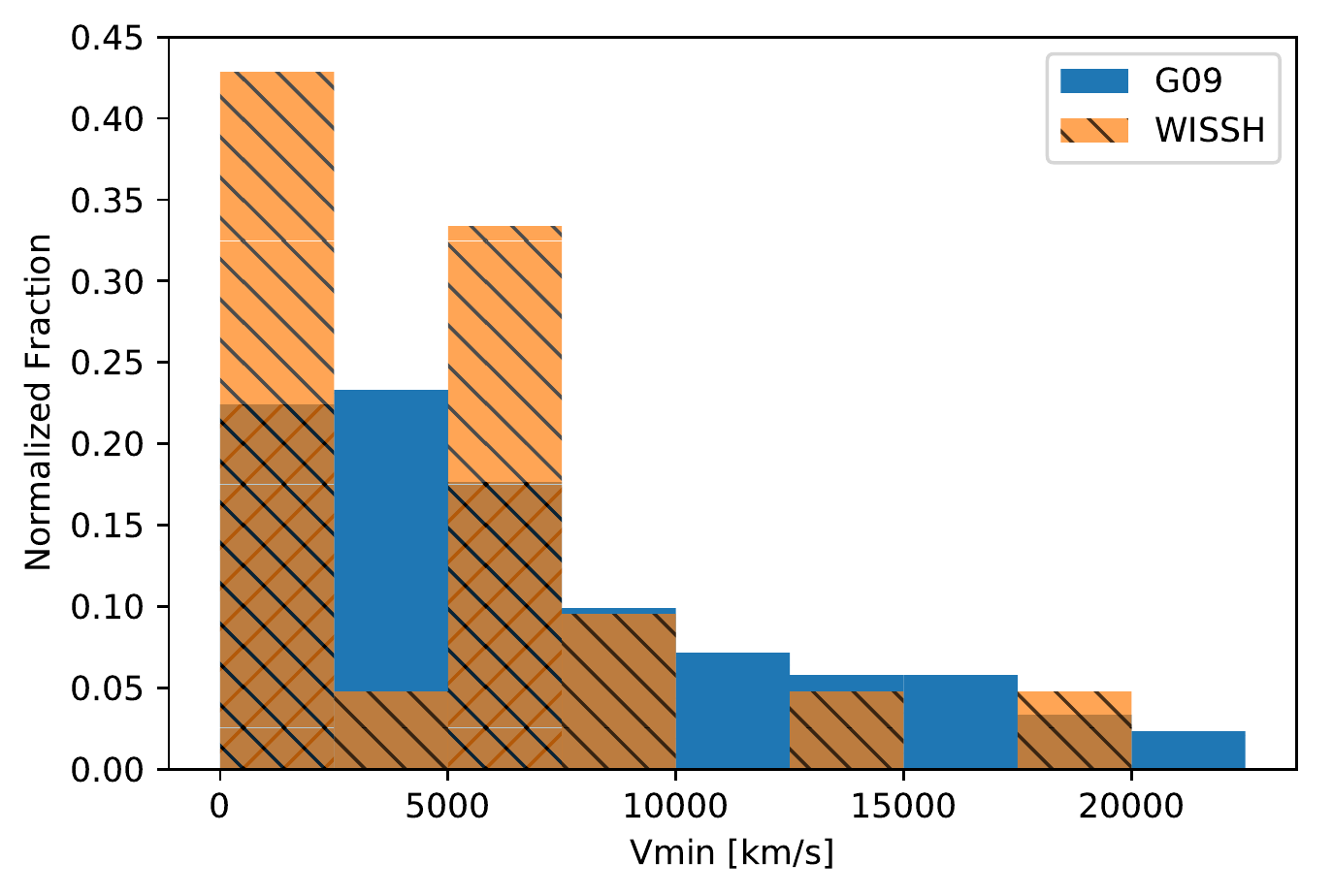}
\label{vmin}
\end{subfigure}

\begin{subfigure}[b]{1.0\linewidth}
\includegraphics[width=\linewidth]{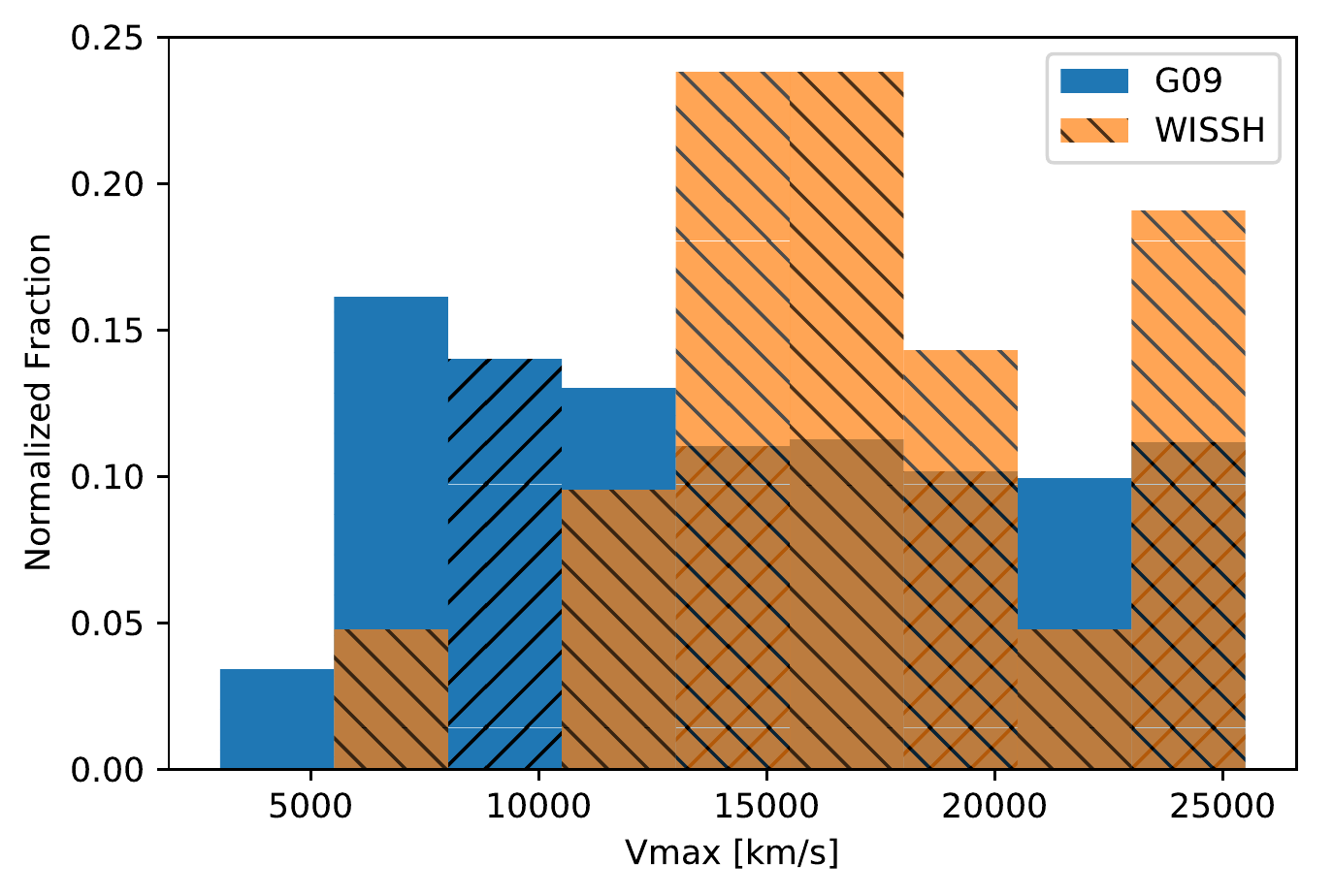}
\label{vmax}
\end{subfigure}

\caption{Comparison between BI$>$0 BAL QSOs $v_{min}$ and $v_{max}$ in WISSH and G09 BAL QSOs catalogue.}\label{VMIN+VMAX}
\end{figure}



\begin{figure}[t]
\centering
\includegraphics[width=1.0\linewidth]{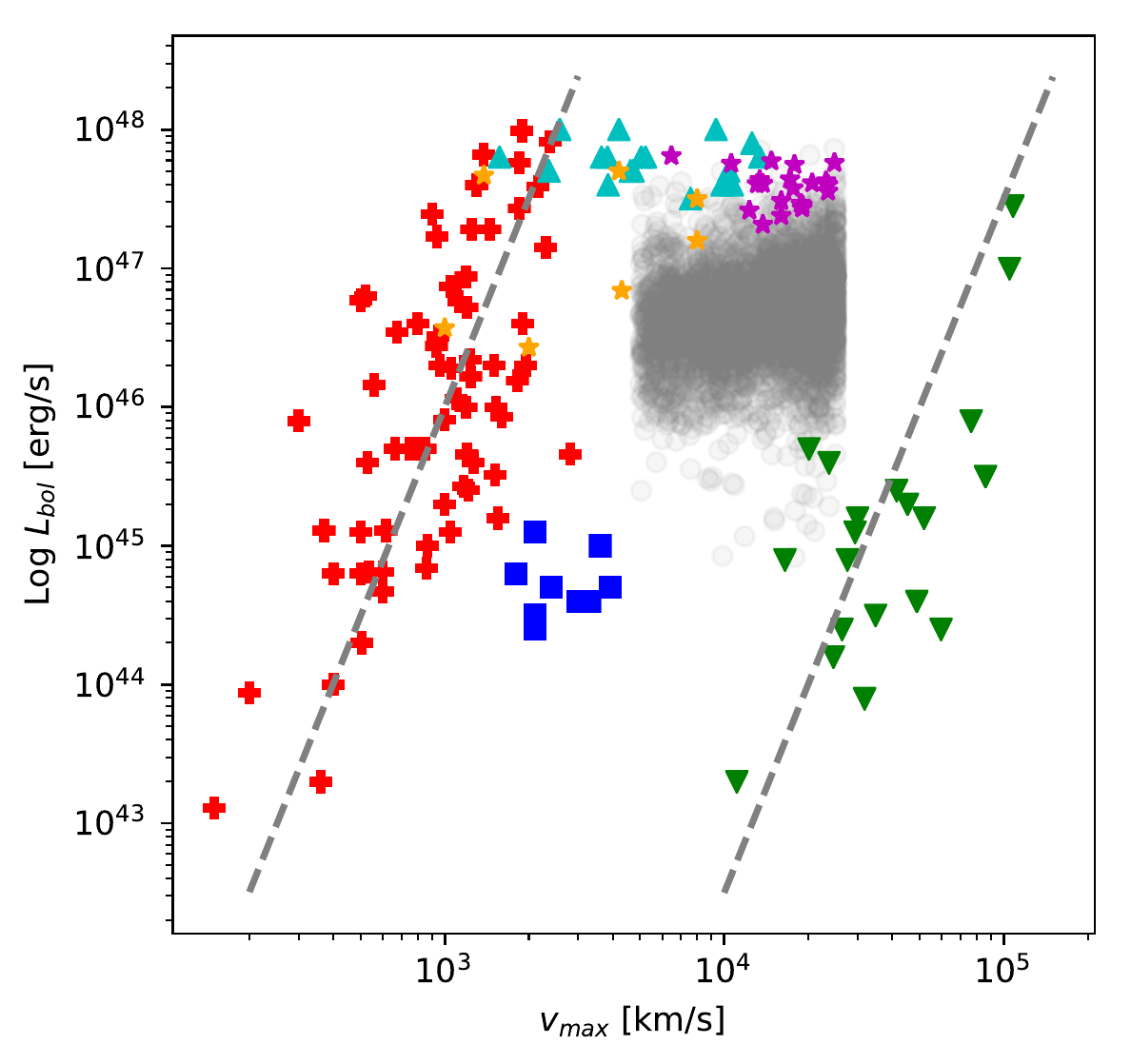}
\caption{Bolometric luminosities versus maximum wind velocity for the objects in \cite{2017A&A...601A.143F} (red crosses: molecular and ionized winds; blue squares: warm absorbers; green downwards triangles: UFOs; orange stars: BALs), for WISSH BI$>$0 BAL QSOs (purple stars for the 21 \civ BALs), and for \civ winds from \cite{2018A&A...617A..81V} (cyan upwards triangles).  The region occupied by the G09 objects with BI$>$0 is in faded gray (note that the artificial $v_{max}$ boundaries are a consequence of the BI selection). }
\label{Lbol_vs_vmax}
\end{figure}


\begin{figure*}[t]
\centering
\includegraphics[width=1.0\linewidth]{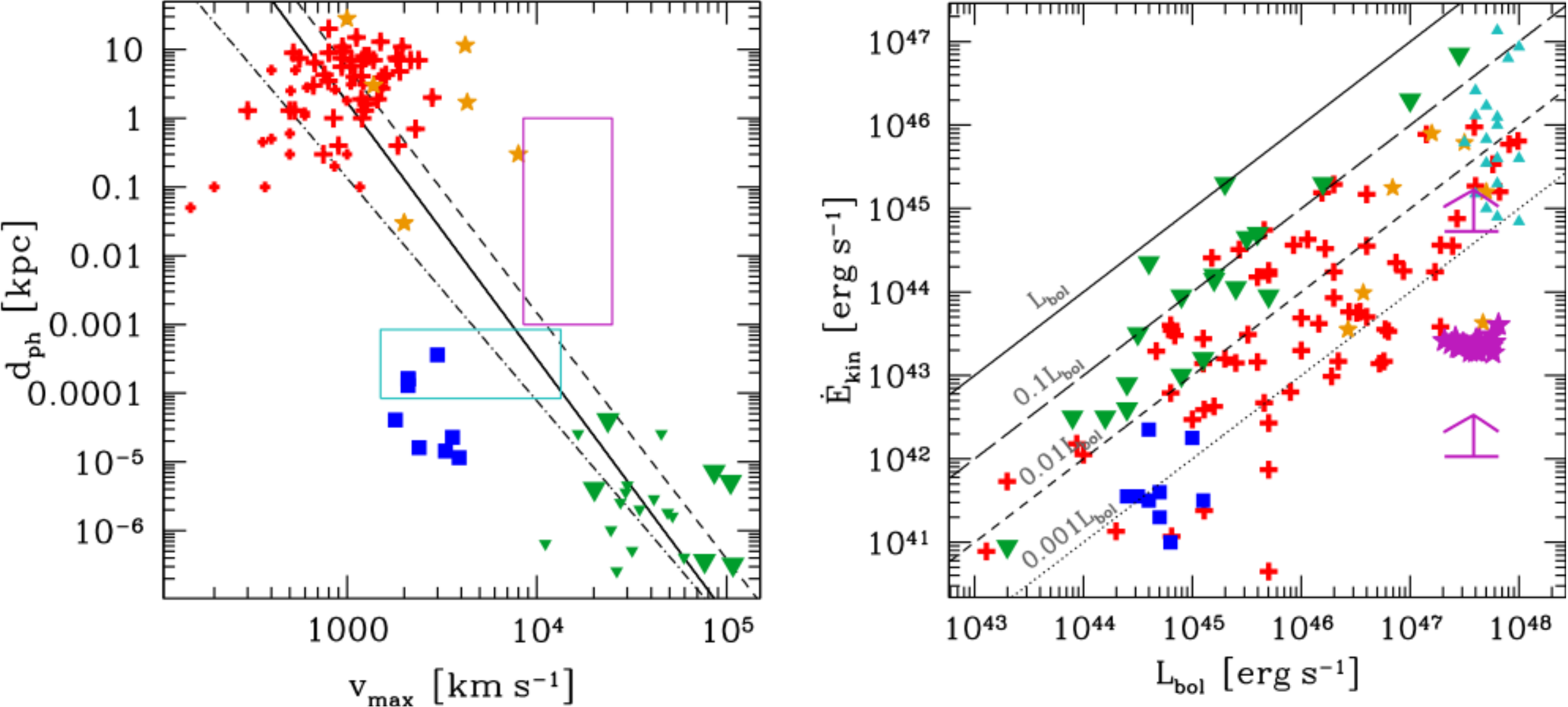}
\caption{Left panel: physical distance ($d_{ph}$) of the high velocity outflow component as a function of $v_{max}$ for the same sources reported in Fig.~\ref{Lbol_vs_vmax} and belonging to the \cite{2017A&A...601A.143F} sample. Large and small symbols for molecular/ionized kpc-scale outflows (red crosses) and UFOs (green triangles) represent sources with bolometric output higher and lower than $\rm \log (L_{bol}/erg\,s^{-1})=45.5$. Solid, dot-dashed and dashed lines report a linear relations connecting the average points of kpc-scale outflows and UFOs for all sources, and sources fainter and brighter than $\rm \log (L_{bol}/erg\,s^{-1})=45.5$, respectively. The position occupied by WISSH BALs and \civ broad emission line winds is reported by magenta and cyan rectangular regions, respectively (see text for details). Right panel: kinetic power as a function of bolometric luminosity for the sources reported in Fig.~\ref{Lbol_vs_vmax}. The two lower limits represent our uncertainty on the BAL column density (see text for details).}
\label{d_vs_vmax}
\end{figure*}


\subsection{Velocity distribution and potential for feedback of BAL QSOs}

It is instructive to compare the distributions for the minimum and maximum velocities of the \civ BAL troughs ($v_{min}$ and $v_{max}$, respectively), for all the BI$>$0 objects from both samples. We define as $v_{min}$ and $v_{max}$ the velocities estimated from the minimum and maximum wavelengths of the BAL troughs in the region between \civ and \siiv (values are given in Tab. \ref{sources}). Normalized distributions for the two samples are plotted in Fig. \ref{VMIN+VMAX}. The $v_{max}$ distribution seems clearly different beetwen the two samples with the WISSH QSOs exhibiting a higher fraction at high velocities. A KS test gives a $p<0.015$ for the $v_{max}$ distribution from WISSH to be drawn from the same parent distribution as G09. This suggests that on average hyperluminous QSOs host the more extreme manifestations of the BAL phenomenon. On the other hand for the $v_{min}$ distributions the test suggests no significant difference between the two samples ($p>0.15$). 

It is tempting to associate $v_{max}$ to the terminal velocity reached at larger distances \citep{2007ApJ...665..990G}. This may imply a more efficient acceleration mechanism, for the most luminous QSOs, which pushes most of the winds up to $\gtrsim$15,000 $\rm km\,s^{-1}$, i.e. at the edge between the \civ and the \siiv emission lines used for the BI classification. As discussed above, the two UBOs show  absorption even further blue-ward of the \siiv line: this further hints to a larger velocity distribution for the WISSH QSOs with even larger $v_{max}$ values. These results show how QSOs with extreme luminosities such as the WISSH ones are able to accelerate BAL winds to larger velocities compared to AGN in lower luminosity regimes. Indeed, G09 found a highly significant correlation between $L_{\lambda2500}$ and $v_{max}$, pointing towards a mainly radiative acceleration for the BAL winds, as suggested by previous authors (\citealt{1995ApJ...454L.105M, 2002ApJ...569..641L}).

In Fig.~\ref{Lbol_vs_vmax} we report $L_{bol}$ as a function of $v_{max}$ for a compilation of AGN-driven winds/outflows measured in different gas phases (molecular, ionized, warm absorbers, and UFOs). Data for molecular and ionized winds (red crosses), warm absorbers (blue squares), and UFOs (green downward triangles), have been collected by \cite{2017A&A...601A.143F} with the specific aim of reporting properties of outflows with available estimates or limits on the physical distance of the high velocity gas from the central engine. Fig. \ref{Lbol_vs_vmax} also includes the WISSH BI$>$0 BAL QSOs together with the incomplete but representative list of BAL QSOs provided by \cite{2017A&A...601A.143F}. Notice that the reported $L_{bol}$ for the WISSH BAL QSOs are estimated throught spectral energy distribution (SED) fitting (Duras et al. in prep.), while a good part of the sources reported in \cite{2017A&A...601A.143F} have $L_{bol}$ estimated also through bolometric correction factors. \cite{2017A&A...601A.143F} found a log-linear relation between $L_{bol}$ and $v_{max}$ for both kpc-scale molecular/ionized and nuclear sub-pc relativistic winds with a similar slope of $\sim5$. The dependence is clearly visible in the plot in which more luminous sources tend to exhibit faster and nuclear/kpc-scale winds. 
In the plot we also report the 21 \civ WISSH BAL QSOs with BI$>$0 as purple stars. For comparison the position occupied by G09 BI$>$0 BALs is  also reported as faded grey region.  Furthermore, we include broad \civ line nuclear winds detected as highly blueshifted emission lines \citep{2018A&A...617A..81V}.
We mention that the log-linear relations found by \cite{2017A&A...601A.143F} normalized to the average $v_{max}$ of the WISSH BALs nicely cover the region occupied by the G09 BALs, hinting to a possible similar relation for this wind phase despite the large spread of values in the G09 BALs.

In Fig.~\ref{d_vs_vmax}, left panel, we report the outflow physical distance ($d_{ph}$) measured for the high velocity gas as a function of $v_{max}$ for kpc-scale outflows, WAs, BALs and UFOs, as reported in  \cite{2017A&A...601A.143F}. 
We also include the locus occupied by BALs (purple rectangle) and \civ broad-emission-line nuclear winds (cyan rectangle) reported for the WISSH quasars. As these winds do not have an actual estimate of the distance, we adopted typical distances reported in literature studies. For WISSH BALs we adopt a range of 1 pc to 1 kpc (see e.g. \citealt{2018ApJ...857...60A, 2017MNRAS.468.4539M, 2018ApJ...866....7L}). For \civ broad-emission-line winds we adopt a range of $\sim100-1000$ light-days typically found for luminous QSOs (e.g. $L_{bol}\approx10^{47-48}\rm erg/s$) by reverberation mapping studies (e.g. \citealt{2007ApJ...659..997K, 2016A&A...587A..43S}). 
Notice that overall $d_{ph}$ and $v_{max}$ seem to anti-correlate with nuclear winds exhibiting faster winds than galaxy-scale outflows. We report in figure a linear relation joining the average 
points for kpc-scale outflow (red crosses) and UFOs (green triangles). There is an indication that this relation seems to be dependent on luminosity. Indeed dot-dashed and dashed lines report the relation for sources fainter and brighter than $\log (L_{bol}/\rm erg ~s^{-1})=45.5$. Notice that given (i) the lack of uncertainties in the data included in the fit and (ii) the incompleteness of the sample used, this relation must not be considered as universal but just as an broad-brush indication of the inverse relation. We find relations with slopes $-3.6$ and $-3.2$ and y-intercept $11.5$ and $8.9$ for bright and faint sources,  respectively. 

We use this relation between $v_{max}$ and $d_{ph}$  in order to estimate the maximum kinetic power $\dot{E}_{kin}$ associated to BAL outflows in WISSH QSOs. We may infer $\dot{E}_{kin}$  by dividing the kinetic energy $E_{kin}=0.5 M\,v^2$ by a characteristic flow time given by $d_{ph}/v_{max}$. The kinetic energy is estimated as in \cite{2019MNRAS.483.1808H}. Their calculation  assumes a spherical shell expanding at a certain velocity ($v$)  with a given covering factor ($Q$) and thickness small compared to its radial distance ($R$). Under these assumptions $E_{kin}$ is expressed by the following formula: 
\begin{equation}
    \begin{split}
    E_{kin}^{max}\approx 1.7\times10^{54}\Big(\frac{Q}{0.28}\Big)
    \Big(\frac{N_{\rm H}}{5\times10^{22}\, cm^{-2}}\Big)
    \Big(\frac{R}{1\, pc}\Big)^2 \\ \times~ \Big(\frac{v}{8000\, km~s^{-1}}\Big)^2 erg.
    \end{split}
\end{equation}
For our calculation we assume $v=v_{max}$, $R=d_{ph}$ and Q=0.28 which is the average value derived from the literature \citep{2006ApJS..165....1T, 2008MNRAS.386.1426K, 2009ApJ...692..758G, 2011MNRAS.410..860A}. Since we do not have information on the column densities, and given that their estimate is beyond the aim of this work, we adopt a value of $ N_{\rm H}=2.2\times10^{21}\, \rm cm^{-2}$. The latter is the logarithmic mean value between 
 of $ N_{\rm H}=10^{20}\, \rm cm^{-2}$ and $ N_{\rm H}=5\times10^{22}\, \rm cm^{-2}$ which are minimum column densities derived from doubly and triply ionized species \cite[][]{2018ApJ...857...60A} and P\,{\sc v} troughs \cite[][]{2019MNRAS.483.1808H}, respectively. 
In Fig.~\ref{d_vs_vmax}, right panel, we report the $\dot{E}_{kin}$ estimated for the WISSH BALs as a function of $L_{bol}$ compared to the same values reported for other winds/outflows by \cite{2017A&A...601A.143F} and \cite{2018A&A...617A..81V}. 
We report as lower limits, the values of $\dot{E}_{kin}$ estimated at the minimum $N_{\rm H}$ reported for low column density \cite[][]{2018ApJ...857...60A} and high column density \cite[i.e. from P\,{\sc v} troughs,][]{2018ApJ...857...60A} winds.
Notice that the densest winds can reach values of kinetic power which are of the order of $\gtrsim0.1$\% of the bolometric luminosity. Even higher values are expected if the luminosity dependence of the $d_{ph}-v_{max}$ relation holds at highest luminosities. Notice that we are making the assumption that the most of the outflowing mass is carried at $v_{max}$. If instead we adopt $v_{min}$ as representative phase for carrying the majority of the kinetic power of the wind, then by adopting the lowest value of $d_{ph}=1$~pc, we obtain a lower limit on $\dot{E}_{kin}$ which is one order of magnitude lower.  Bearing in mind  the crude assumptions in this estimate, this is an indication that some of these winds are in principle able to transport a kinetic power sufficient for generating a significant feedback contribution on the host \citep{2005Natur.433..604D,2010MNRAS.401....7H}.


\subsection{Multiple AGN winds in WISSH quasars}

In \cite{2017A&A...598A.122B} and \cite{2018A&A...617A..81V} we revealed the presence of kpc- and BLR-scale winds in WISSH objects by detecting a broad/blue-shifted \oiii\, or \civ emission lines in the rest-frame, optical, and UV spectra of a randomly-selected sub-sample of 18 WISSH QSOs. Among them, seven objects (namely 1157+27, 1326-00, 1421+46, 1422+44, 1538+08, 1549+12 and 2123-00) exhibit absorption features in their SDSS spectra (see Table \ref{sources}). Specifically, they all show an AI$_{1000}>0$ km/s), and for four sources we derive a positive BI. Vietri et al. (2018) reported the existence of two sub-populations of WISSH QSOs based on their emission line properties. The first one (consisting of six sources, dubbed \oiii) is characterized by a broad \oiii\, emission line, a Rest-frame Equivalent Width (REW) of the \civ emission REW$\rm_{CIV}$ $\approx$ 20-40 \AA\  and a profile of the \civ emission line blueshifted of $<$ 2,000 km/s. The second one (dubbed \weak\ sources, and representing two-thirds of the LBT/LUCI WISSH sample) exhibits weak/absent \oiii\ emission, and a highly-blue-shifted (2,000 -- 8,000 km/s) \civ emission line with REW $<$ 20 \AA. 

\cite{2018A&A...617A..81V} interpreted the dichotomy observed in WISSH QSOs in terms of a combination of huge ionizing flux (leading to overionization of the NLR gas and, hence, a decrease of the \oiii\, emission, e.g. \citealt{2014Natur.513..210S}) and inclination. In particular, for \weak\ QSOs the accretion disk is seen face-on, while \oiii\ ones should be viewed at larger inclination angles, i.e. $\theta \sim$25--70 degrees. Interestingly, intermediate inclinations ($\sim$25--40 degrees) for BAL QSOs have been also suggested by \cite{2000ApJ...545...63E}. 

Among the seven QSOs with  AI $>$ 0 km/s, three belong to the \oiii\ subclass, while four are \weak\ objects. Two out of three \oiii\ QSOs are classified as BAL according to BI, while the remaining one (1326-00) has AI$_{1000}$ $\simgt$ 2,000 km/s. This is quite interesting and suggests that  the simultaneous presence of BAL and kpc-scale outflows may be common at the highest AGN luminosities. On the contrary, only one \weak\ QSO (1157+27) shows a BI  $>$ 0 km/s, and all the remaining ones only have modest (\simlt 300 km/s) AI. Future follow-up studies of the \oiii\ emission in a larger sample of luminous BAL QSOs will be crucial to support this orientation scenario for the simultaneous detection of NLR and BAL outflows.


\section{Radio properties of WISSH BAL QSOs}

We cross-correlated the BAL QSOs list from WISSH with the FIRST catalogue (\citealt{1995ApJ...450..559B}), using a 5" matching radius centered on the optical coordinates, in order to determine the fraction of WISSH BAL QSOs with a radio counterpart. Eight over 38 ($\sim$21\%) have a counterpart in FIRST, namely 0747+27, 1025+24, 1130+07, 1204+33, 1237+06, 1422+44, 1513+08, and 1549+12. As a whole, 20 over 86 objects ($\sim$23\%) in WISSH have a radio counterpart. 

\begin{figure}
\centering
\includegraphics[width=\linewidth]{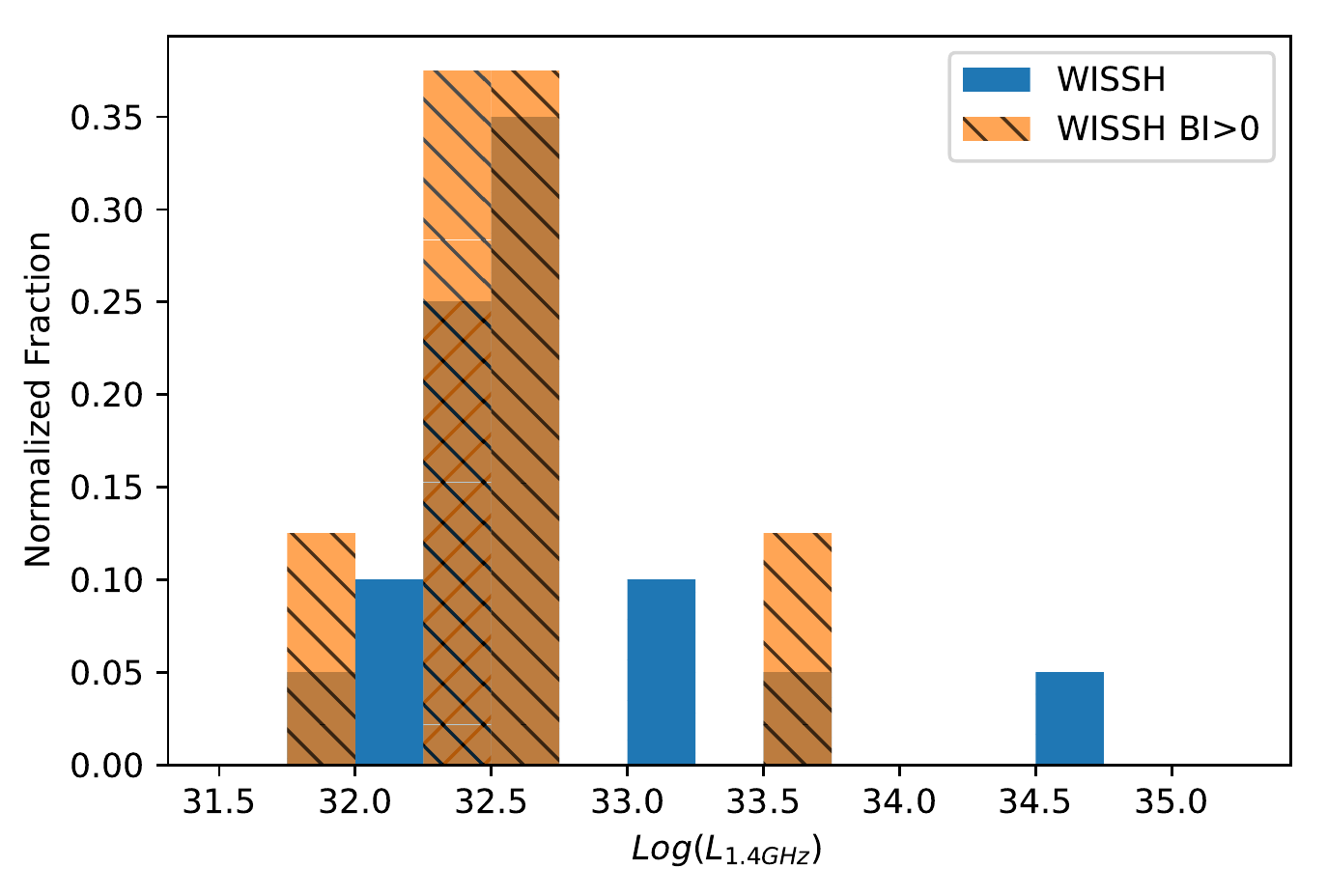}
\caption{Normalized distributions of $L_{1.4\,\rm{GHz}}$ for all WISSH objects with FIRST radio counterpart (blue) and for BAL QSOs in WISSH with FIRST radio counterpart (orange).}\label{L14}
\end{figure}

\cite{2008ApJ...687..859S} studied the dependence of the BAL QSOs fraction among the radio population as a function of specific luminosity at 1.4 GHz ($L_{1.4\,GHz}$, from FIRST). They found that it can drop from $\sim$20\% to $\sim$8\% when moving from an $L_{1.4\,GHz}\sim10^{32}$ to $\sim10^{36}$ erg/s/Hz for BI-selected BAL QSOs, while from $\sim$45\% to $\sim$20\% for the AI-selected ones. No entirely satisfactory physical model was found for such a behavior, both the evolutionary and geometrical one not reflecting the complex phenomenology of these objects. The $L_{1.4\,GHz}$ range for the WISSH sample is $8\times10^{31}-2\times10^{35}$ erg/s/Hz (see Fig. \ref{L14}), similar to the one explored by those authors. This was calculated from the FIRST flux density as in equation 1 of \cite{2008ApJ...687..859S}, but assuming a spectral index $\alpha$=0 since both inverted and steep spectral indices have been found in samples of radio-loud BAL QSOs (\citealt{2008MNRAS.388.1853M, 2011ApJ...743...71D, 2012A&A...542A..13B}) - although, as discussed in \cite{2008ApJ...687..859S}, the estimate has only a weak dependence on $\alpha$.
Despite the poor statistic - only 8 among the 20 WISSH object with a radio counterpart are BAL QSOs - the fraction we find at luminosities $L_{1.4\,GHz}<10^{33}$ erg/s/Hz is $\sim$47\% when considering objects with AI$>$0, and $\sim$20\% when considering the ones with BI$>$0, in agreement with the trend found by those authors.

From a morphological point of view, images from FIRST show a compact morphology for all sources, corresponding to an upper limit for linear sizes of $\sim$40 kpc at the mean redshift of the BAL QSOs subsample in WISSH ($z$=3.2). This is in agreement with linear sizes from previous studies at arcsec angular resolution (\citealt{2012A&A...542A..13B}), resulting in a compact radio morphology, $<$40 kpc, for more than 90\% of objects at a similar redshift range.  

The radio properties presented here do not suggest a behavior different from previous samples, despite the extreme luminosity regime in WISSH. The coupling between the BAL-producing winds and the jet seems to have trends similar to the ones presented in previous works from the literature, meaning that the higher probability for WISSH objects to launch winds does not imply different jet formation rates or strength.\\


\section{Conclusions}

We have analysed the properties of the BAL QSOs fraction in the WISSH sample, composed by hyper-luminous type-1 AGN, and compared to those of BAL QSOs from the literature. Results can be summarized as follows:

\begin{itemize}

\item Adopting the standard BAL indices, we found in WISSH 38 objects with an AI$>$0 (44$\pm$7\%), 32 with AI$_{1000}>$0 (37$\pm$6\%), and 21 with BI$>$0 (24$\pm$5\%).

\item The fraction of objects with a \civ BI$>$0 (24\%) is almost two times higher than what found in G09. Also, the BI distributions for the two samples are distinct, with typically higher values for the WISSH sample. The LoBAL fraction among BAL QSOs in WISSH is $\sim$26\%, which is also larger than what found at similar luminosities in previous works. The higher $L_{\rm bol}$ of the WISSH objects likely  favours the acceleration of BAL outflows (in both the high- and low-ionization gas components), indicating  that they are likely radiatively driven.

\item The maximum velocities of the \civ BAL outflows in WISSH have a different distribution from those in G09, with WISSH QSOs exhibiting larger values. This lends further support to the theoretical predictions suggesting that, via radiative pressure, hyper-luminous QSOs are able to launch the most extreme winds \citep{2019arXiv190407341G}.

\item We find two QSOs exhibiting BAL features with velocities of $\sim$0.15c, by assuming that absorption blue-wards of the \siiv emission line is associated with \civ. Although the statistics is very limited (i.e. 2 with respect to the 21 objects showing a BI$>$0), this seems to lend further support to a scenario where hyper-luminous QSOs are able to accelerate the most powerful outflows (e.g. \citealt{2002ApJ...569..641L, 2017A&A...601A.143F})

\item We estimate the possible range of kinetic power associated to BAL outflows in WISSH which can reach values above 0.1\% of $\rm L_{bol}$. 
This indicates that, especially for the highest column density and fastest winds, BAL outflows in hyperluminous QSOs are potentially able to provide an efficient feedback onto the host galaxy interstellar medium.

\item About 20\% of BI-selected BAL QSOs from WISSH shows a radio counterpart in FIRST (1.4 GHz), this fraction being compatible with the one found in G09 (23\%). It also confirms that the WISSH sample shows the same anti-correlation between $L_{1.4 GHz}$ and BAL QSOs abundance found by \cite{2008ApJ...687..859S}, not indicating any particular dependence of the BAL QSOs radio properties on the $L_{\rm bol}$.

\end{itemize}

Future works based on much larger samples of hyper-luminous QSOs are needed to  shed more light on the crucial role of $L_{bol}$ in increasing the fraction and the power of BAL outflows in the QSO population.


\begin{acknowledgements}
We thank the anonymous referee for the constructive comments and suggestions.
GB acknowledges financial support under the INTEGRAL ASI-INAF agreement 2013-025-R.1. 
EP acknowledges financial support from the Italian Space Agency (ASI) under the contracts ASI-INAF I/037/12/0 and ASI-INAF n.2017-14-H.0. 
LZ acknowledges financial support under ASI/INAF contract I/037/12/0.
FT acknowledges support by the Programma per Giovani Ricercatori - anno 2014 Rita Levi Montalcini. This research project was supported by the DFG Cluster of Excellence ‘Origin and Structure of the Universe’ (www.universe-cluster.de). 

Funding for the Sloan Digital Sky Survey IV has been provided by the Alfred P. Sloan Foundation, the U.S. Department of Energy Office of Science, and the Participating Institutions. SDSS-IV acknowledges
support and resources from the Center for High-Performance Computing at
the University of Utah. The SDSS web site is www.sdss.org.

SDSS-IV is managed by the Astrophysical Research Consortium for the 
Participating Institutions of the SDSS Collaboration including the 
Brazilian Participation Group, the Carnegie Institution for Science, 
Carnegie Mellon University, the Chilean Participation Group, the French Participation Group, Harvard-Smithsonian Center for Astrophysics, 
Instituto de Astrof\'isica de Canarias, The Johns Hopkins University, 
Kavli Institute for the Physics and Mathematics of the Universe (IPMU) / 
University of Tokyo, the Korean Participation Group, Lawrence Berkeley National Laboratory, 
Leibniz Institut f\"ur Astrophysik Potsdam (AIP),  
Max-Planck-Institut f\"ur Astronomie (MPIA Heidelberg), 
Max-Planck-Institut f\"ur Astrophysik (MPA Garching), 
Max-Planck-Institut f\"ur Extraterrestrische Physik (MPE), 
National Astronomical Observatories of China, New Mexico State University, 
New York University, University of Notre Dame, 
Observat\'ario Nacional / MCTI, The Ohio State University, 
Pennsylvania State University, Shanghai Astronomical Observatory, 
United Kingdom Participation Group,
Universidad Nacional Aut\'onoma de M\'exico, University of Arizona, 
University of Colorado Boulder, University of Oxford, University of Portsmouth, 
University of Utah, University of Virginia, University of Washington, University of Wisconsin, 
Vanderbilt University, and Yale University.
\end{acknowledgements}


   \bibliographystyle{aa} 
   \bibliography{WISSH} 


\begin{appendix}

\section{BAL QSOs spectra}

 We report here the SDSS DR12 spectra of the 38 BAL QSOs presented in this work. For each object, we show the spectrum, the combined continuum and emission lines fit performed in {\tt IRAF}, the \siiv and \civ peaks position as calculated from redshift (dashed gray lines), and the residuals used for the BAL indices integrals calculation (from 0 to 25,000 km/s), where the absorption below 90\% of the continuum is highlighted in orange. We also mark the $v_{min}$ and $v_{max}$ estimates for each object as dashed orange lines.


\begin{figure*}
\centering
\includegraphics[width=0.95\linewidth]{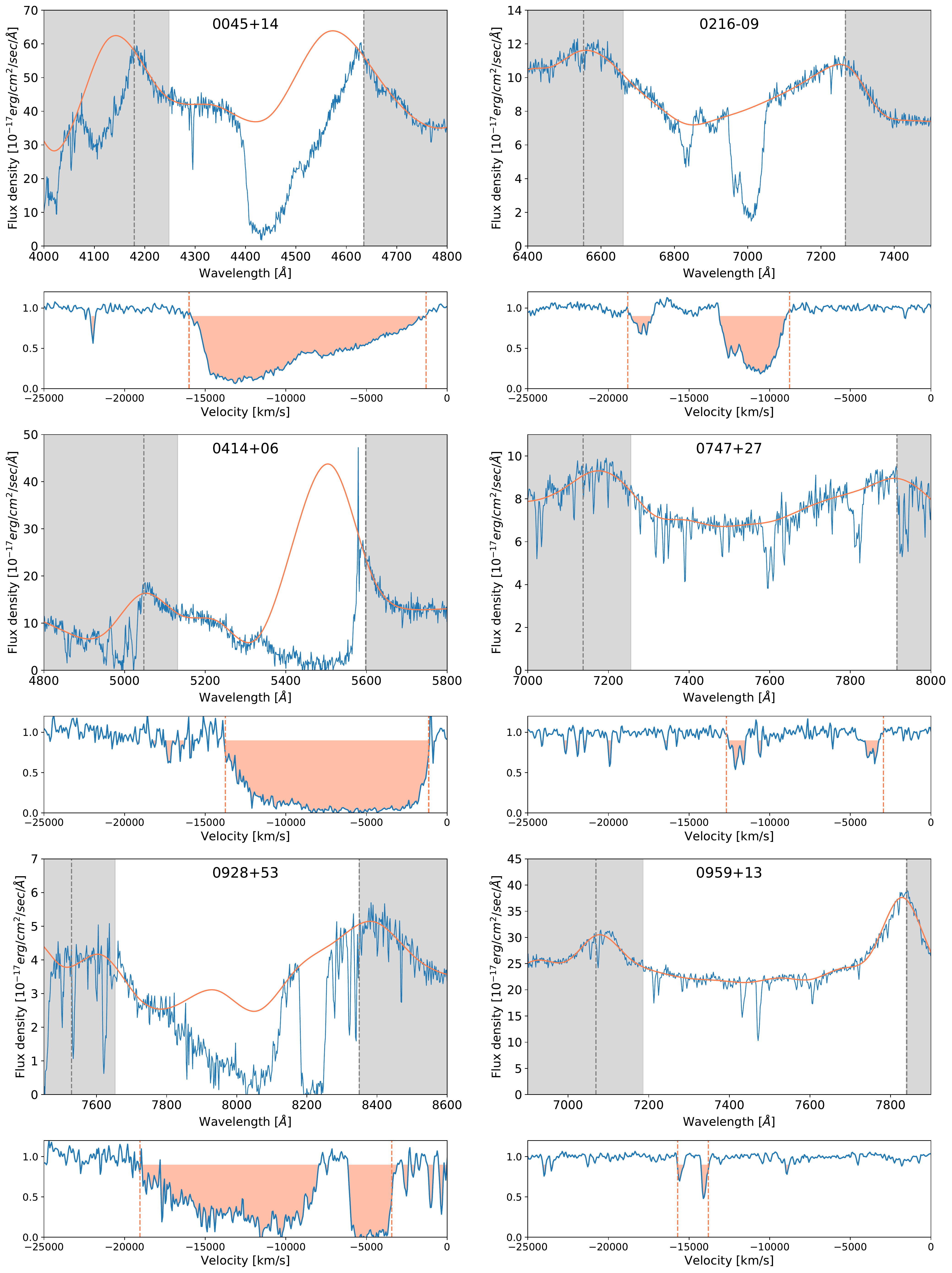}
\caption{SDSS DR12 spectra of the 38 BAL QSOs presented here. Top panel: spectrum (blue line) and fit performed in {\tt IRAF} (orange line); dashed lines indicate the position of the \siiv and \civ peaks as calculated from redshift, while the white area the wavelengths interval between 0 and $-$25,000 km/s. Bottom panel: residuals between the \civ peak and $-$25,000 km/s, with absorption below 90\% of the continuum highlighted in orange; dashed lines indicate the minimum and maximum velocity estimated for the BAL outflow.}\label{new_BALs}
\end{figure*}

\begin{figure*}
\ContinuedFloat
\centering
\includegraphics[width=0.95\linewidth]{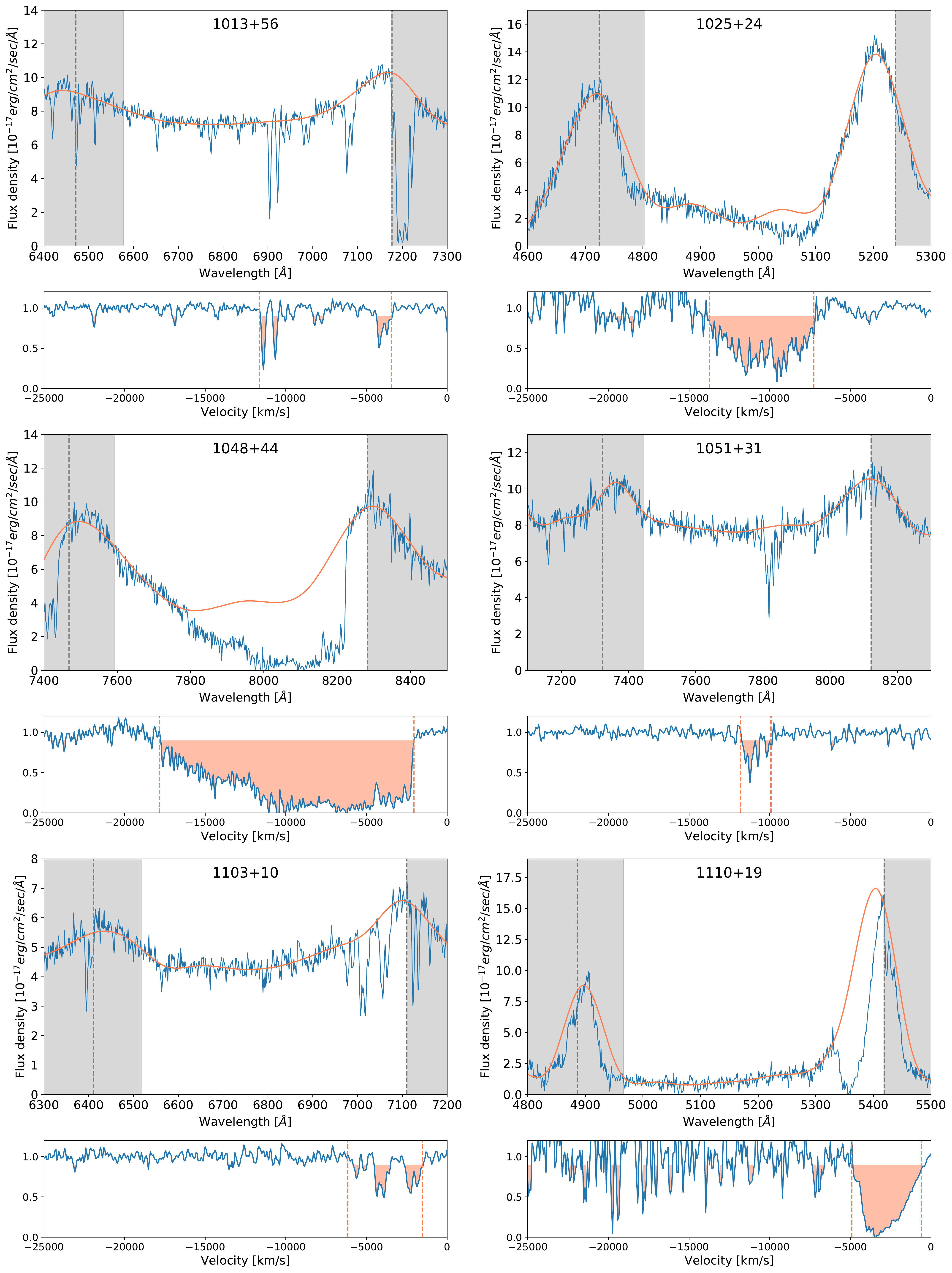}
\caption{Continued}\label{new_BALs}
\end{figure*}

\begin{figure*}
\ContinuedFloat
\centering
\includegraphics[width=0.95\linewidth]{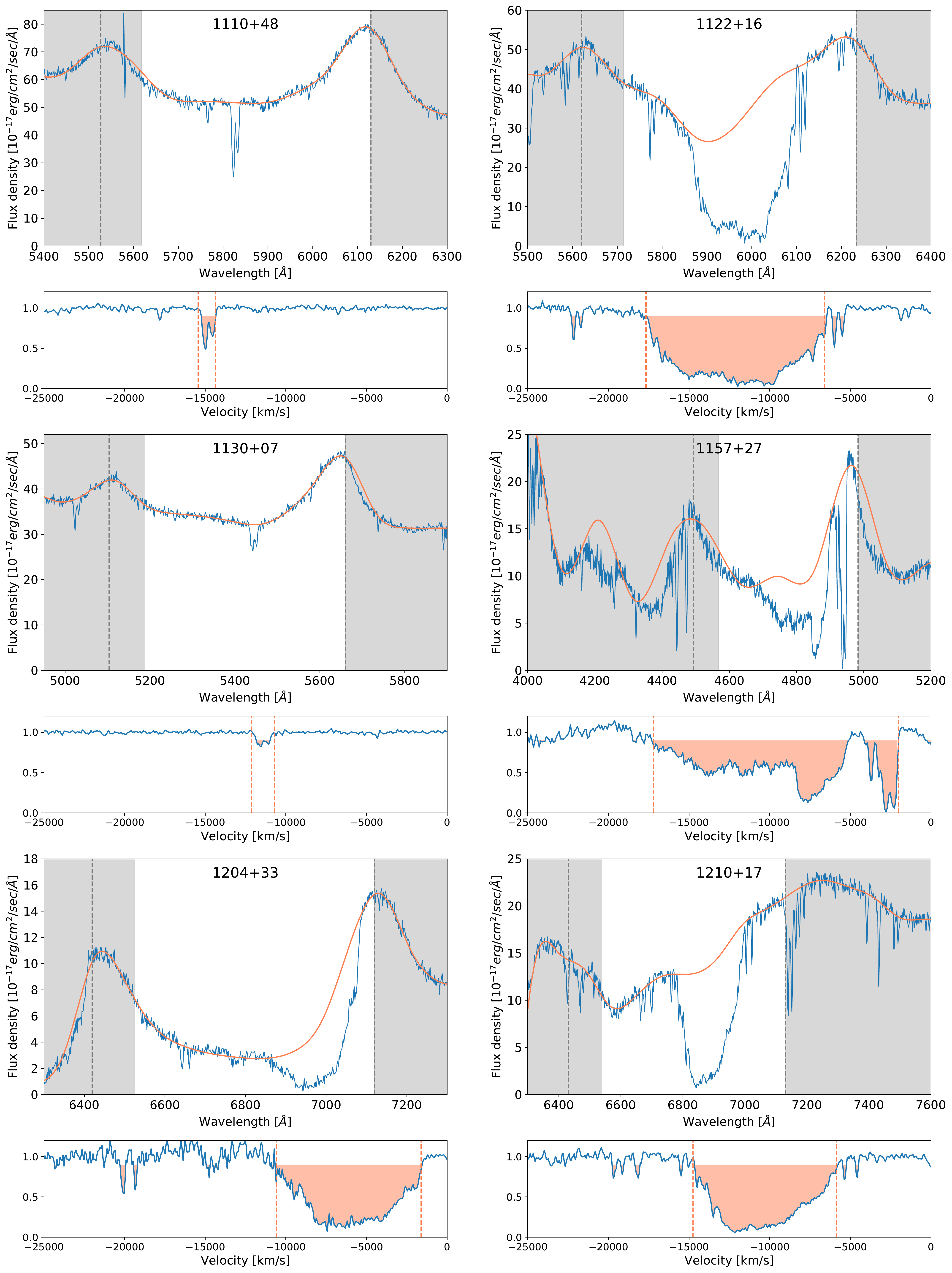}
\caption{Continued}\label{new_BALs}
\end{figure*}

\begin{figure*}
\ContinuedFloat
\centering
\includegraphics[width=0.95\linewidth]{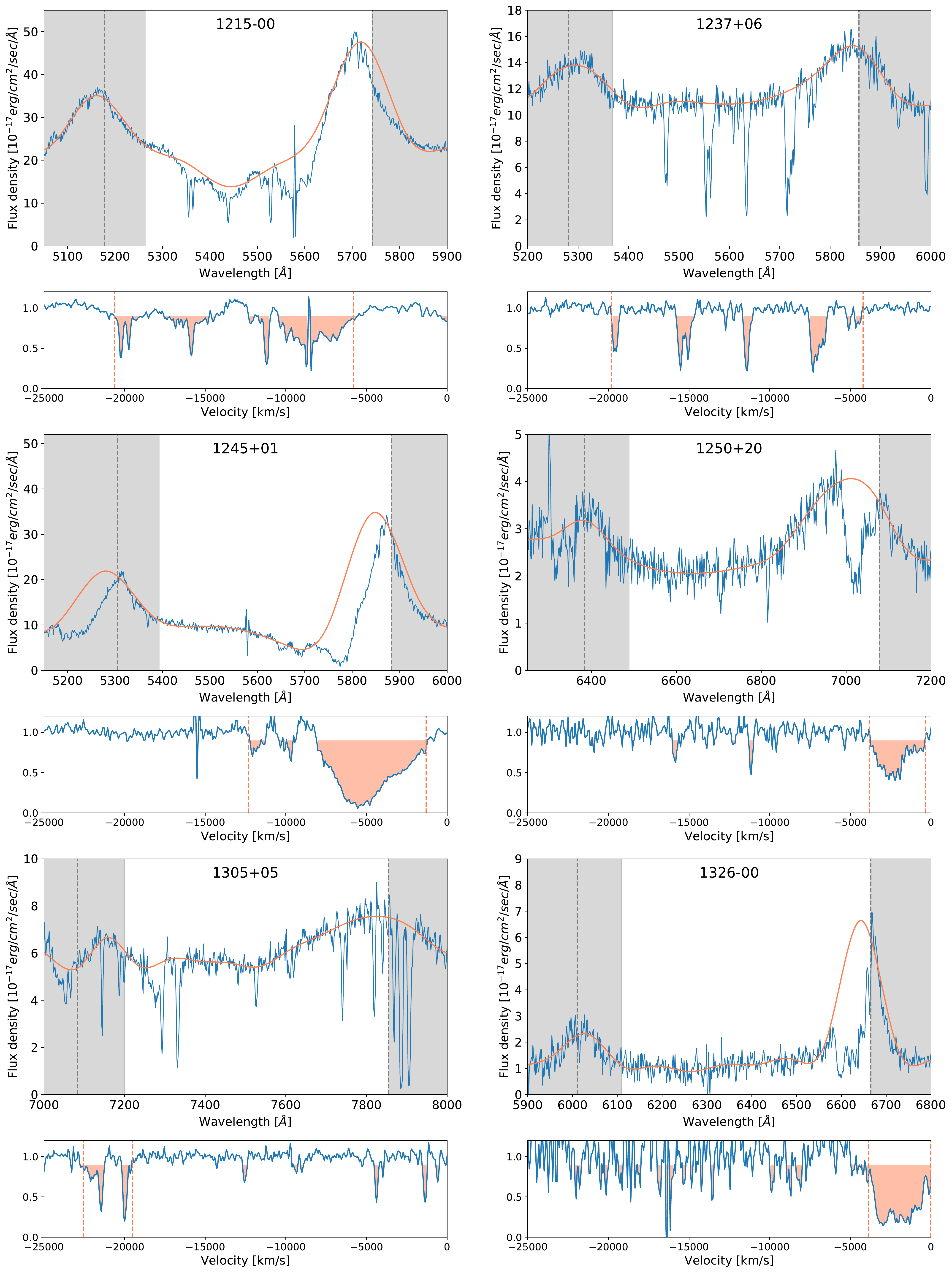}
\caption{Continued}\label{new_BALs}
\end{figure*}

\begin{figure*}
\ContinuedFloat
\centering
\includegraphics[width=0.95\linewidth]{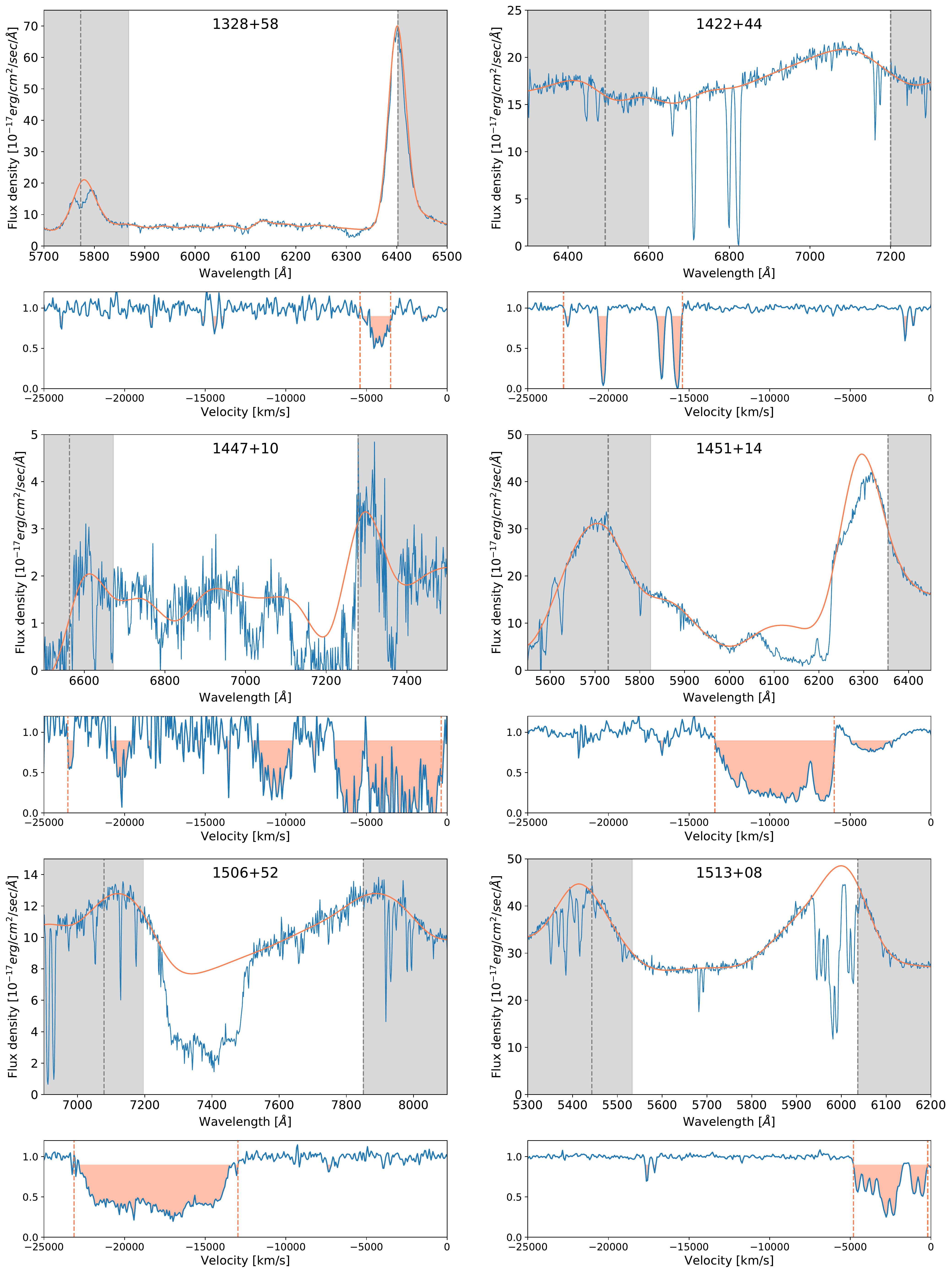}
\caption{Continued}\label{new_BALs}
\end{figure*}

\begin{figure*}
\ContinuedFloat
\centering
\includegraphics[width=0.95\linewidth]{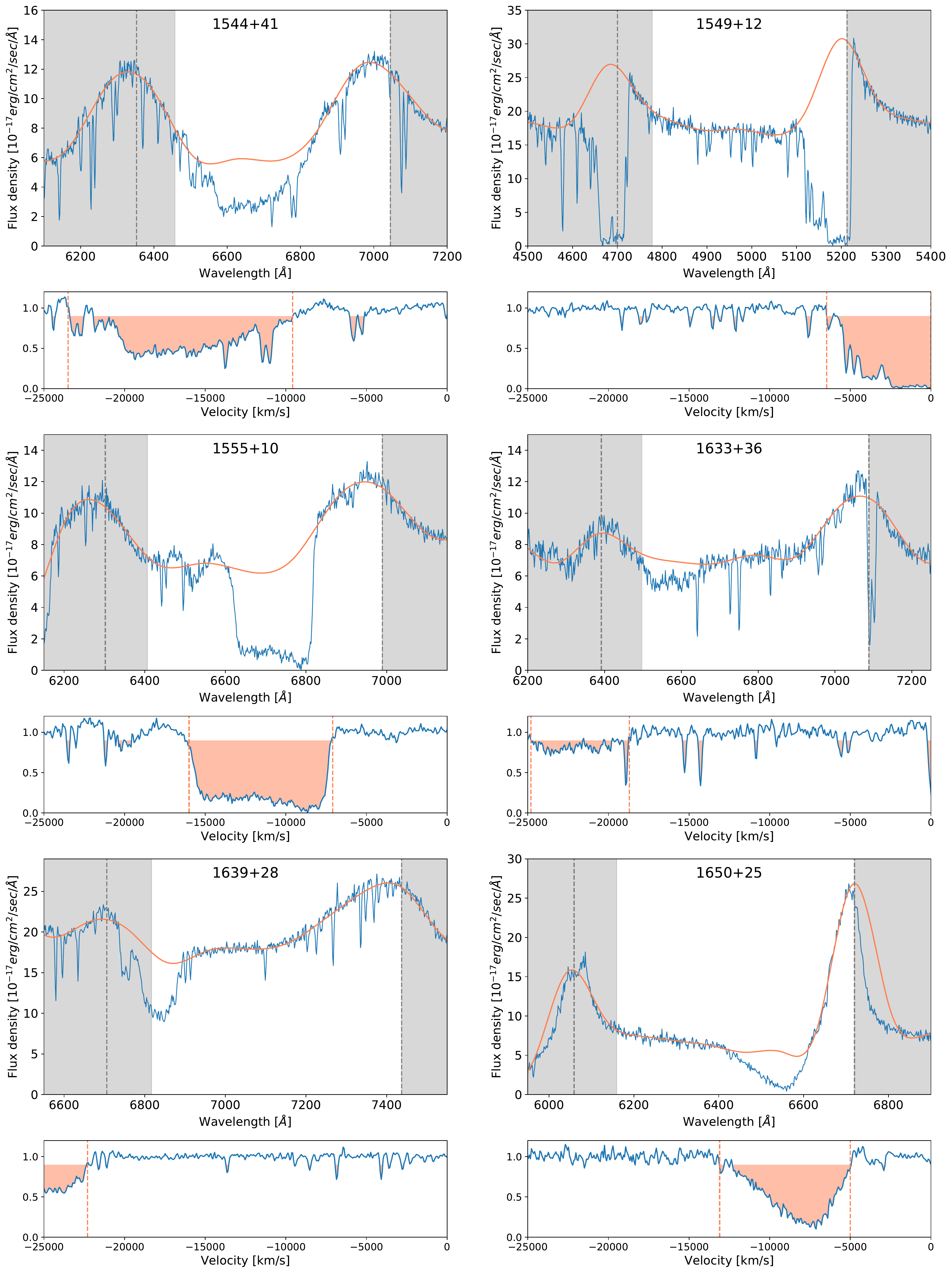}
\caption{Continued}\label{new_BALs}
\end{figure*}

\begin{figure*}
\ContinuedFloat
\centering
\includegraphics[width=0.95\linewidth]{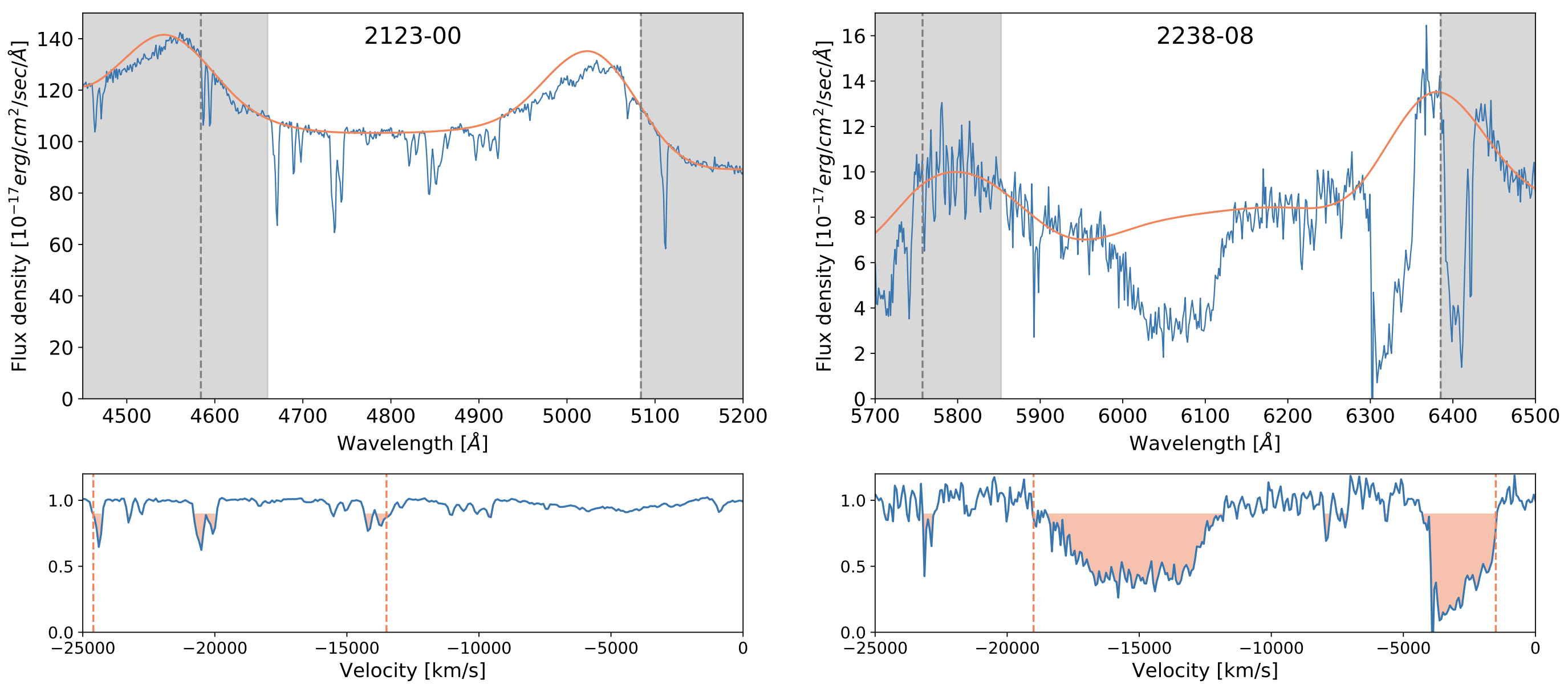}
\caption{Continued}\label{new_BALs}
\end{figure*}

\end{appendix}


\end{document}